\documentclass[fleqn,usenatbib]{mnras}

\usepackage{newtxtext,newtxmath}

\usepackage[T1]{fontenc}
\usepackage{ae,aecompl}

\usepackage{graphicx}	
\usepackage{amsmath}	

\newcommand{\shark}{\textsc{Shark}}
\newcommand{\cf}{$_\mathrm{CF00}$}
\newcommand{\trr}{$_\mathrm{T20-RR14}$}
\newcommand{\sting}{\textsc{Stingray}}
\newcommand{\prospect}{\textsc{ProSpect}}
\newcommand{\viper}{\textsc{Viperfish}}
\newcommand{\gc}{G$^3$C}
\newcommand{\rap}{$r_\mathrm{ap}$}
\newcommand{\giap}{${g-i}_\mathrm{ap}$}

\title[Synthetic galaxy colours]{From rest-frame luminosity functions to observer-frame colour distributions: tackling the next challenge in cosmological simulations}

\author[Bravo et al.]{
Mat\'ias Bravo$^1$\thanks{E-mail:matias.bravo@icrar.org},
Claudia del P. Lagos$^{1,2,3}$,
Aaron S.~G. Robotham$^{1,2}$,
\newauthor{
Sabine Bellstedt$^1$,
Danail Obreschkow$^{1,2}$}
\\
$^{1}$International Centre for Radio Astronomy Research (ICRAR), M468, University of Western Australia, 35 Stirling Hwy, Crawley, \\WA 6009, Australia.\\
$^{2}$ARC Centre of Excellence for All Sky Astrophysics in 3 Dimensions (ASTRO 3D).\\
$^{3}$Cosmic Dawn Center (DAWN). 
}

\date{Accepted XXX. Received YYY; in original form ZZZ}

\pubyear{2019}

\begin{document}
\label{firstpage}
\pagerange{\pageref{firstpage}--\pageref{lastpage}}
\maketitle

\begin{abstract}
Galaxy spectral energy distributions (SEDs) remain among the most challenging yet informative quantities to reproduce in simulations due to the large and complex mixture of physical processes that shape the radiation output of a galaxy.
With the increasing number of surveys utilising broadband colours as part of their target selection criteria, the production of realistic SEDs in simulations is necessary for assisting in survey design and interpretation of observations.
The recent success in reproducing the observed luminosity functions (LF) from far-UV to far-IR, using the state-of-the-art semi-analytic model \shark\ and the SED generator \prospect, represents a critical step towards better galaxy colour predictions.
We show that with \shark\ and \prospect\ we can closely reproduce the optical colour distributions observed in the panchromatic GAMA survey.
The treatment of feedback, star formation, central-satellite interactions and radiation re-processing by dust are critical for this achievement.
The first three processes create a bimodal distribution, while dust attenuation defines the location and shape of the blue and red populations.
While a naive comparison between observation and simulations displays the known issue of over-quenching of satellite galaxies, the introduction of empirically-motivated observational errors and classification from the same group finder used in GAMA greatly reduces this tension.
The introduction of random re-assignment of $\sim 15\%$ of centrals/satellites as satellites/centrals on the simulation classification closely resembles the outcome of the group finder, providing a computationally less intensive method to compare simulations with observations.
\end{abstract}



\section{Introduction}\label{sec:intro}

The colours of a galaxy, namely the ratio of the observed flux at two different wavelength bands of a galaxy, are among the most direct observables. They are, however, the end product of the interplay of many complex physical processes and hence challenging to decipher.
Observed colours combine information from star formation rates (SFR), stellar populations, metal and dust production and distribution, with none of these being simple processes from a physics perspective \citep[see the review by][]{conroy2013}.
The advent of the Sloan Digital Sky Survey \citep[SDSS;][]{york2000} cemented our understanding of the optical colour distribution of galaxies, convincingly proving that it is bi-modal \citep[e.g.,][]{strateva2001,hogg2002}, and that colours depend on both the galaxy's stellar mass and its environment \citep[e.g.,][]{baldry2006,peng2010}.

Beyond the wealth of information contained in colour distributions, their use also has become more prevalent in the past decade as part of extragalactic survey designs.
Surveys such as WiggleZ \citep{drinkwater2010}, the Baryon Oscillation Spectroscopic Survey \citep[BOSS;][]{dawson2013}, and the Dark Energy Spectroscopic Instrument \citep[DESI;][]{desi2016} survey directly employed colour selections for their target selection.
Colour information can also be used indirectly, for example on the planned Wide-Area VISTA Extragalactic Survey \citep[WAVES;][]{driver2019}, where the aim is to conduct target selection based on photometric redshifts.
Understanding the biases introduced by colour-derived selection is critical for the science cases of such surveys.

Currently, there is a wide variety of methods available for the production of synthetic galaxy catalogues, ranging from purely empirically-driven to fully modelling all relevant physical processes in galaxy formation and evolution \citep[for a recent overview see][]{wechsler2018}.
While empirical methods by construction reproduce the observables on which they are based, the predicting power of the physically-driven methods is required when testing our understanding of the physical processes that shape galaxies. Physical models are also fundamental when the galaxy properties targeted by future surveys is beyond current observations.
Despite their vital role, the reproduction of observed colour distributions has remained a challenge for galaxy formation simulations.

\citet{guo2016} analysed the Evolution and Assembly of GaLaxies and their Environments \citep[EAGLE][]{schaye2015} hydrodynamical simulations in tandem with two Semi-Analytic Models (SAM) run on the dark matter (DM)-only version of EAGLE, \textsc{L-galaxies} and GALFORM, and found no good agreement between the simulations and observations of the fraction of passive galaxies, even when this is computed from SFRs rather than colours \citep[see also][]{ayromlou2020}.
Unlike colours, SFRs are a direct output from all physically-driven models.
The tension between simulations and observations is greatest for low-mass satellites \citep[e.g.,][]{font2008,guo2016,cucciati2017}, with these galaxies predicted to be more quenched than observed.
\citet{font2008} found that the GALFORM \citep[e.g.,][]{cole2000,baugh2005,lagos2012,lacey2016}  SAM suffered from this issue, and showed that changing the gas stripping of galaxies when becoming a satellite, from instantaneous to gradual \citep[following][]{mccarthy2008}, greatly reduced the tension with observations.

While that prescription of stripping has not been adopted in the most recent version of GALFORM \citep{lacey2016}, similar models of non-instantaneous gas stripping have been adopted on other SAMs.
Semi-Analytic Galaxy Evolution \citep[\textsc{SAGE}; e.g.,][]{croton2006,croton2016} adopts gradual gas stripping but fails to reproduce the observed passive fractions measured from specific star formation rates \citep[sSFR;][]{croton2016}.
\textsc{DarkSAGE} \citep{stevens2016,stevens2017} and Semi-Analytic Galaxies \citep[\textsc{SAG}; e.g.,][]{lagos2008,cora2018} also include forms of gradual stripping and manage to produce a better agreement for the sSFR-derived passive fraction of satellites, at least at $z=0$ \citep{stevens2017,cora2018}.
Interestingly also Galaxy Evolution and Assembly \citep[\textsc{GAEA}; e.g.,][]{hirschmann2016,delucia2019} can reproduce the observed sSFR-derived passive fraction of galaxies while using instantaneous stripping.
This tells us that what matters is not a single physical process but instead the interplay between all the baryon physics included in galaxy formation models.
In addition, the exact passive fraction prediction depends on the way this is defined (e.g. via colours or SFRs).
Some models use some version of passive fraction in their calibration and hence can reproduce them by construction \citep[e.g., \textsc{L-Galaxies}][]{guo2011,henriques2015}.
The drawback of the latter approach is that the reproduction of a given colour-defined passive fraction does not guarantee other definitions to be well reproduced.

The fact that different models adopting different physical descriptions of a range of baryon physics achieve reasonable  sSFR-derived passive fractions is a symptom of a broader problem.
\citet{mitchell2018} and \citet{lagos2018} showed that galaxy formation models using vastly different approaches, and more importantly, different physical descriptions for any one physical process, produce similar stellar mass growth rates and SFR evolution. This results from the degeneracy between different physical models and parameters included in galaxy formation simulations. 
Hence, we require more complex tests to distinguish between models, and in this paper we argue that colour distributions as a function of stellar mass and cosmic time provide such test.
From the available theoretical models, \shark\ \citep[hereafter L18]{lagos2018}\defcitealias{lagos2018}{L18} is among the most promising ones to achieve a good match to observed colours.
While it has not been tested for sSFR-derived passive fractions to this point, \citet[hereafter L19]{lagos2019}\defcitealias{lagos2019}{L19} showed that \shark\ is capable of reproducing observed luminosity functions and number counts across a wide range of bands, from the far-ultraviolet (far-UV) to the far-infrared (far-IR) and from $z=0$ to $z=10$.
This was achieved using a combination of \shark\ with Spectral Energy Distribution (SED) fitting/generation software \prospect\ \citep{robotham2020} and the parametrised \citet{charlot2000} attenuation curve proposed by \citet{trayford2020} using the Radiative Transfer post-processing of the EAGLE simulations.
The results presented in \citetalias{lagos2019} have encouraged us to study the colour distributions that the combination of \shark\ and \prospect\ predict, as the successful reproduction of luminosity functions across several bands should imply that colours are reasonable.
However, as we want to test colours as a function of mass and time, it is not straightforward  that \shark, or for that matter any model that produces reasonable luminosity functions, is able to do this.

This work is structured as follows.
In Section \ref{sec:mocks} we introduce the GAMA catalogues used in this work, together with how we construct synthetic galaxy catalogues to reproduce GAMA.
We compare the observed and synthetic colour distributions from our catalogues, and the blue and passive fractions of both in Section \ref{sec:colours}.
We discuss our findings in Section \ref{sec:discuss}, and summarise our work in Section \ref{sec:summary}.


\section{Building synthetic observations}\label{sec:mocks}
With its unique mix of high redshift completeness ($95\%$) down to a magnitude of $r=19.5$ and availability of a wide range off broad-band photometry  (from the far-UV to far-IR), the galaxy catalogues made from the equatorial fields of the Galaxy And Mass Assembly \citep[GAMA,][]{driver2011,liske2015} survey are the prime data set for environmental studies of galaxies.
For this work, we have combined the latest version of the GAMA Galaxy Group Catalogue \citep[\gc,][hereafter R11]{robotham2011}\footnote{\url{http://www.gama-survey.org/dr3/schema/dmu.php?id=21}}\defcitealias{robotham2011}{R11}, the newly made \textsc{KidsVikingGAMA} photometric catalogue from far-UV to far-IR \citep{bellstedt2020a}, and an extension of the new catalogue of physical properties \citep{bellstedt2020b} produced using the fitting mode of the software \prospect.

To construct the \gc\ catalogue,  \citetalias{robotham2011} used a Friend-of-Friends (FoF) group finder, with separate linking lengths for the projected and radial directions.
The algorithm was built to take into account both the redshift completeness of the survey near each galaxy and the average density of galaxies given both the GAMA survey LF and magnitude limit.
The free parameters of the group finder were calibrated using $9$ synthetic light-cones (LC) made with the Millennium DM-only simulation and the GALFORM \citep{bower2006} SAM, with the $r$ magnitudes of the galaxies re-adjusted to follow the GAMA redshift-dependant LF and selection function.
The calibration aimed for both high bijectivity, having most of the groups in the synthetic LCs recovered and few spurious detections, and high purity, with most of the galaxies in recovered groups being part of the matching group in the mock lightcone.
The group finding has only been released for the equatorial GAMA regions, so these are the only GAMA regions suitable to study galaxy environment and central-satellite populations.
For this reason we will refer to the GAMA equatorial fields simply as GAMA for the rest of this work.

\citet{bellstedt2020a} have recently produced an all-new photometric catalogue for GAMA (equatorial fields and G23), \textsc{KidsVikingGAMA}, using the software \textsc{ProFound}\footnote{\url{https://github.com/asgr/ProFound}} \citep{robotham2018} for the source detection, on imaging from the Galaxy Evolution Explorer \citep[GALEX,][]{martin2005} space telescope for the FUV-NUV bands, VLT Survey Telescope (VST) Kilo-Deegre Survey \citep[VST,][]{arnaboldi2007} for the $u$-$g$-$r$-$i$ bands, the Visible and Infrared Survey Telescope for Astronomy \citep[VISTA,][]{sutherland2015} VISTA Kilo-Degree Infrared Galaxy Survey \citep[VIKING,][]{arnaboldi2007} for the $Z$-$Y$-$J$-$H$-$Ks$ bands, Wide-field Infrared Survey Explorer \citep[WISE,][]{wright2010} for the $W1$-$W2$-$W3$-$W4$ bands, and the Herschel Space Observatory \citep{pilbratt2010} for the 100-160-250-350-500 $\mu$m bands.
This new source finding is the reason that the completeness and magnitude value previously mentioned ($95\%$ for $r<19.5$) differ from the literature values of $98\%$ completeness down to $r=19.8$ \citep{bellstedt2020b}.

\textsc{ProFound} was specifically designed to overcome existing problems on common source finding software, like the use of spherical or elliptical apertures instead of isophotes and double-counting of flux in the case of overlapping apertures.
Source detection was conducted on a stack of the $r$ and $Z$-band images, and the photometry extracted for the $u$-$g$-$r$-$i$-$Z$-$Y$-$J$-$H$-$Ks$-$W1$-$W2$ bands.
As seen in figure 14 of \citet{bellstedt2020a}, the new photometry is consistent with the previous set \citep[LAMBDAR,][]{wright2016}, save for the FUV and NUV bands.
\citet{bellstedt2020a} show that the FUV-NUV is slightly bluer in \prospect\ compared to LAMBDAR, but better behaved, with fewer gross outliers through a wide choice of colours.
For the mid- and far-IR, a different process was used for the photometry, as galaxies are usually not resolved in those bands.
However, in this work we do not use the mid- and far-IR bands to compare with our simulations, focusing instead on the optical regime where both stellar emission and dust attenuation play a significant role, and therefore we refer the reader to \citet{bellstedt2020a} for details on this process.

\prospect\footnote{\url{https://github.com/asgr/ProSpect}} \citep{robotham2020} is a low-level Spectral Energy Distribution (SED) generator, with several of the design choices influenced by existing spectral fitting codes like \textsc{MAGPHYS} \citep{dacunha2008} and \textsc{CIGALE} \citep{noll2009,boquien2019}.
It combines either the \textsc{GALEXev} \citep{bruzual2003} or \textsc{E-MILES} \citep{vazdekis2016} Stellar Population Synthesis (SSP) libraries with the \citet{charlot2000} multi-component dust attenuation model and the \citet{dale2014} dust re-emission templates, under the assumption of a \citet{chabrier2003} Initial Mass Function (IMF); identical to that used in \shark. Following \citetalias{lagos2019}, we have chosen to use the \textsc{GALEXev} SSP due to the wider wavelength coverage that it provides \citep[see figure 21 of][]{robotham2020}.

In the fitting mode, \prospect\ offers a wide variety of choices of functional forms for the characterisation of the star formation and metallicity histories (SFH and ZH, respectively).
To fit the photometry from the \textsc{KidsVikingGAMA} catalogue, \citet{bellstedt2020b}:
\begin{itemize}
    \item choose the \texttt{massfunc\_snorm\_trunc} functional form for the SFH, with $m_\mathrm{SFR}$, $m_\mathrm{mpeak}$, $m_\mathrm{mperiod}$, and $m_\mathrm{mskew}$ as free parameters to fit.
    This parametrisation represents a skewed normal distribution, where the skewness, width, peak position and peak height are all free parameters, with an additional constraint that the star formation is anchored at 0 at the start of the Universe.
    \item parametrise the ZH using \texttt{Zfunc\_massmap\_box}, which maps the build-up of stellar mass via the fitted SFH onto the build-up of metals, using a closed-box model.
    They include $Z\mathrm{final}$ as a free parameter, such that the final metallicity of the ZH is free.
    \item Leave the $\tau_\mathrm{birth}$, $\tau_\mathrm{screen}$, $\alpha_\mathrm{birth}$ and $\alpha_\mathrm{screen}$ dust parameters as free parameters within the fitting.
    \item Fix pow$_\mathrm{birth}$ and pow$_\mathrm{screen}$ to the default \prospect\ values.
    \item Set the maximum age of the Universe, the look-back time at which star formation is allowed to begin, to 13.8 Gyr.
\end{itemize}
To fit the parameters, a Covariance Matrix Adaptation genetic algorithm is applied to get an initial guess of the parameters, and then a Component-wise Hit And Run Metropolis Markov-Chain Monte Carlo algorithm is used with $10,000$ steps to determine the best-fitting SFH, ZH and dust parameters. 
The use of \prospect\ has slightly raised the measured stellar masses at low redshift ($\sim0.15$ dex at $z<0.1$) compared to the existing GAMA catalogues \citep[see Figure 33 of][]{robotham2020}.


\subsection{Building synthetic universes}

Inspired by the success of \citetalias{lagos2019} in reproducing the observed galaxy LF across a wide choice of filters, we have chosen to use the same models to produce GAMA-like synthetic LCs.
This requires the use of:

\begin{itemize}
\item an $N$-body DM-only simulation to calculate the time-dependant distribution and dynamics of DM, starting from a chosen cosmology and an initial distribution of said matter at an early stage of the universe's evolution, distributed on a spatial box of fixed comoving size (\S\ref{sec:surfs}).
\item an algorithm to group DM particles into halos and sub-halos at every time-step of the simulation, and a tree builder to establish progenitor-descendant links between the halos in the simulation (\S\ref{sec:surfs}).
\item a SAM that populates halos with galaxies, and evolves them based on a set of equations that model the baryonic processes that shape the evolution of galaxy properties (\S\ref{sec:shark}).
\item an LC builder, to generate a synthetic distribution of galaxies on a given sky footprint, with more distant galaxies [from the observer] being at a higher redshift/look-back time (\S\ref{sec:lc_making}).
\item an SED generator, that uses properties such as the SFH and ZH of galaxies to calculate their panchromatic emission (\S\ref{sec:prospect}).
\end{itemize}

\begin{figure}
    \centering
    \includegraphics[width=\linewidth]{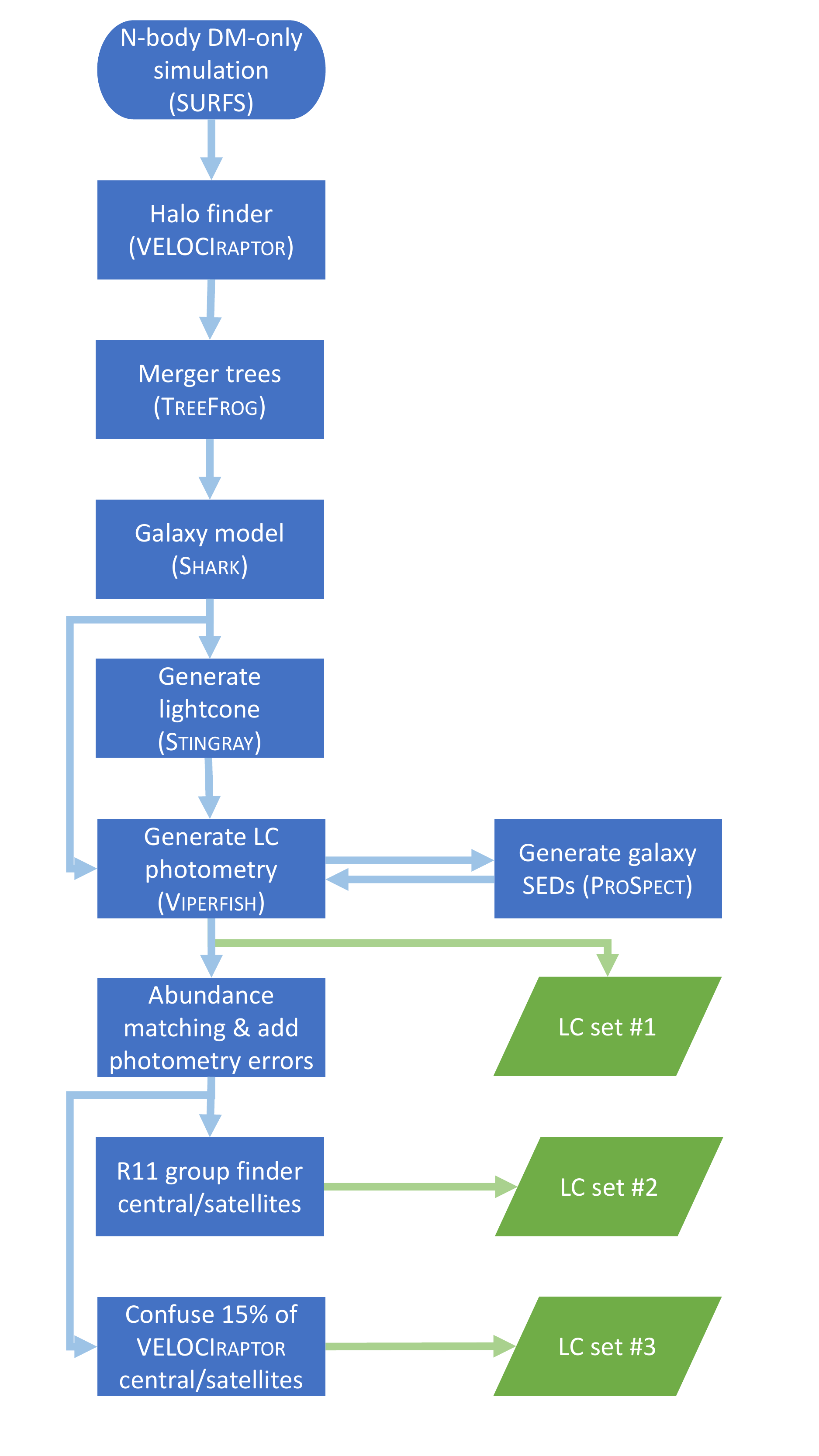}
    \caption{Flow chart that summarises the process used in this work to generate the three synthetic LCs we compare to GAMA throughout this work. The $N$-body DM-only simulation, halo finder and tree builder are described in \S\ref{sec:surfs}; the SAM used in \S\ref{sec:shark}; the lightcone builder in \S\ref{sec:lc_making}; the SED generation in \S\ref{sec:prospect}; the first set of synthetic LCs in \S\ref{sec:LCv1}; the abundance matching, which involves also the adjustment of stellar masses, addition of photometry errors, use of the \citetalias{robotham2011} group finder, and second set of synthetic LC in \S\ref{sec:LCv2}; and the simple model for central/satellite classification and last set of synthetic LC in \S\ref{sec:LCv3}.}
    \label{fig:flow_chart}
\end{figure}

The outcome of this process is the prediction of the colour distribution of galaxies across cosmic time.
The approach we are taking is certainly less expensive than large cosmological hydro-dynamical simulations, such as EAGLE and Illustris-TNG \citep{pillepich2018}, but retains much of the physical description of galaxy formation at an inexpensive computational cost. 
For this work we have expanded on the method used by \citetalias{lagos2019} to generate three sets of synthetic LCs to model GAMA, each set containing two dust models
Figure \ref{fig:flow_chart} presents a schematic view of this process.
The remainder of this section is dedicated to a detailed description of each step.

\subsubsection{DM simulation, halo catalogue and merger tree}\label{sec:surfs}

For this work, we have used the SURFS suite of DM-only simulations \citep{elahi2018b}, which adopts a $\Lambda$CDM \citep{planck2016xiii} cosmology and span a range of box length of $40-210$ cMpc h$^{-1}$ (cMpc being comoving megaparsec) and particle mass of $4.13\times10^7$ to $5.90\times10^9$, reaching up to 8.5 billion particles.
The chosen cosmology has total matter, baryon, and dark energy densities of $\Omega_m=0.3121$, $\Omega_b=0.0491$, and $\Omega_\Lambda=0.6751$, respectively, a Hubble parameter of H$_0=67.51$ km s$^{-1}$ Mpc$^{-1}$, a scalar spectral index of $n_s=0.9653$, and a present root-mean-square matter fluctuation averaged over a sphere of radius $8$ Mpc h$^{-1}$ of $\sigma_8=0.8150$.
This simulation suite was run with a memory lean version of the \textsc{GADGET2} code on the Magnus supercomputer at the Pawsey Supercomputing Centre.
Following \citetalias{lagos2019} we used the L210N1536 simulation for this work, with a box size of 210 cMpc h$^{-1}$, $1,536^3$ DM particles, a particle mass of $2.21\times10^8$ $M_\odot \mathrm{h}^{-1}$, and a softening length of $4.5$ ckpc h$^{-1}$.
SURFS produced 200 snapshots for each simulation, with a typical time-span between snapshots in the range of $\approx6-80$ Myr.

The halo catalogues for the SURFS suite were constructed using the 6D FoF finder \textsc{VELOCIRaptor} \citep{canas2019,elahi2019a}, and for the halo merger trees \textsc{TreeFrog} \citep{elahi2019b} was used.
These catalogues track a two-level hierarchical structure, distinguishing host halos and sub-halos.
Host halos are constructed using a 3D FoF, while sub-halos are the dynamically distinct structures inside each host halo found using a 6D FoF.
At every snapshot, a halo contains one central sub-halo and $\ge 0$ satellite sub-halos.

The combination of \textsc{VELOCIRaptor}+\textsc{TreeFrog} has been comprehensively tested on the SURFS suite, producing well-behaved trees with robustly reconstructed orbits \citep{poulton2018}, and orbits that reproduce the velocity dispersion versus halo mass inferred in observations \citep{elahi2018a}.
We refer the reader to \citetalias{lagos2018} for more details on the construction of the merger trees and halo catalogues used in this work, and to \citet{poulton2018,elahi2019a,elahi2019b,canas2019} for more details on \textsc{VELOCIRaptor} and \textsc{TreeFrog}.

\subsubsection{Populating the simulation with galaxies} \label{sec:shark}

We have used the open-source \shark\footnote{\url{https://github.com/ICRAR/shark}} SAM, introduced by \citetalias{lagos2018}, to follow the formation and evolution of galaxies in our simulated DM-only universe.
While \citetalias{lagos2018} calibrated the free parameters in \shark\ to only reproduce the $z=0,1,2$ stellar mass functions (SMFs), the $z=0$ black hole–bulge mass relation and the mass–size relation, this model has been shown to match a variety of observational measurements, such as the mass–metallicity relations for gas and stars (\citetalias{lagos2018}), the scatter around the main sequence of star formation in the SFR–stellar mass plane \citep{davies2019b}, the HI mass and velocity width of galaxies observed in the ALFALFA survey \citep{chauhan2019}, AGN LFs both X-rays and radio wavelengths \citep{amarantidis2019}, galaxy LFs and number counts from the far-UV to the near-IR (\citetalias{lagos2019}), and the stellar-gas content scaling relations \citep[\citetalias{lagos2018},][]{hu2020}.
This is achieved by including prescriptions for all the physical processes we think shape the formation and evolution of galaxies.
These processes are: 
\begin{enumerate}
\item collapse and merging of DM halos;
\item phase changes of gas between HII, HI and H$_2$;
\item accretion of gas onto halos, which is modulated by the DM accretion rate;
\item shock heating and radiative cooling of gas inside DM halos, leading to the formation of galactic discs via conservation of specific angular momentum of the cooling gas;
\item star formation in galaxy discs; \label{item:sf_chanel_d}
\item stellar feedback from the evolving stellar populations;
\item chemical enrichment of stars and gas;
\item growth of black holes via gas accretion and merging with other supermassive black holes;
\item heating by active galaxy nuclei (AGN);
\item photo-ionisation of the intergalactic medium and intra-halo medium in low mass halos;
\item galaxy mergers driven by dynamical friction within common DM halos, which can trigger starbursts and the formation and/or growth of spheroids; \label{item:sf_chanel_bm}
\item collapse of globally unstable discs that also lead to starbursts and the formation and/or growth of bulges; \label{item:sf_chanel_bd}
\item environmental processes affecting the gas content of satellite galaxies.\label{item:strip}
\end{enumerate}

\shark\ adopts a universal \citet{chabrier2003} IMF and includes several different models for gas cooling, AGN, stellar and photo-ionisation feedback, and star formation, for which, following \citetalias{lagos2019}, we adopt the default models and parameters presented in \citetalias{lagos2018} (see their table 2).

Built into any SAM, including \shark, is the assumption that galaxies at any given time can be described by two components: a disc and a bulge.
These two are distinguished by the mechanism for their formation, with discs building stellar mass consuming gas accreted from the halo into the galaxy, and bulges built by consuming the gas dumped into it during disc instabilities and galaxy mergers, also accreting the stellar material of the satellite in the latter case.

It is important to note that one of the processes included in \ref{item:strip} is the stripping of gas from galaxies when they become a satellite.
This process is assumed to be instantaneous in \shark, something that is not universally adopted across SAMs, as mentioned in Section \ref{sec:intro}.
In addition to gas stripping, satellite subhalos are also assumed to be cut off from cosmological accretion, which means that even in the absence of gas stripping, gas should eventually exhaust in these subhalos via gas cooling and star formation.

When consuming gas for star formation, stars are formed following the surface density of H$_2$, with bulges being more efficient than discs by a factor of $\eta_\mathrm{burst}$, a free parameter in the model with a default value of 10, which is the value obtained in observations of local and high-redshift starburst galaxies \citep{daddi2010,scoville2016,tacconi2018}.
\citetalias{lagos2018} found the latter to be key in reproducing the cosmic star formation rate density (CSFRD) at $z\gtrsim1.5$.
\citetalias{lagos2018} sets as the default model the pressure relation by \citet{blitz2006} to compute the gas partition into HI and H$_2$ as a function of radius.

When a satellite sub-halo merges into the central sub-halo and is lost in the halo catalogue, the galaxy that the satellite sub-halo hosted is re-assigned as an orphan galaxy (\texttt{type 2}) of the central sub-halo, with a merging timescale calculated following the dynamical friction timescale of \citet{lacey1993}.

\subsubsection{Adding light to the simulated galaxies}\label{sec:prospect}

Among the outputs produced by \shark, a \texttt{star\_formation\_histories} file can be produced at every simulation snapshot\footnote{This is done by setting \texttt{output\_sf\_histories = true} on the \shark\ configuration file.}.
These files contain the SFH and ZH, with an entry for every snapshot from the formation of each galaxy to the current simulation time.
The three channels for star formation, \ref{item:sf_chanel_d}, \ref{item:sf_chanel_bm} and \ref{item:sf_chanel_bd} as described in \S\ref{sec:shark}, are tracked separately.

To produce the SED of each galaxy these files are fed to two packages: \prospect\ and \viper\footnote{\url{https://github.com/asgr/Viperfish}}.
In the generative mode of \prospect, discretely-valued SFH and ZH at the observation snapshot are passed to the package, which first calculates the un-attenuated light emission of each galaxy.
Then it adds the screening and re-emission by dust, with all stars in the galaxy screened by dust from the diffuse interstellar medium (ISM).
Stars younger than 10 Myr are assumed to be inside birth clouds, so their light is first attenuated by dust on the cloud, then by the ISM.
In this work we do not attempt to include the effect of AGN emission in \shark.
This is because the current tracked properties (black hole masses and accretion rates) are insufficient to model the AGN mid-IR  emission and additional modelling of black hole properties would be required.
We leave this for future work.

\viper\ is a light wrapper around \prospect, which reads the SFH/ZH from the \shark\ outputs and the desired SED through target filters and pass those to \prospect.
\prospect\ includes a pre-loaded set of 39 filters from the far-UV (Galex FUV) to millimetre-wavelengths (ALMA band 4), with further 347 EAZY \citep{brammer2008} filters available in a loadable table (also included on the package).
As galaxies are treated in SAMs as comprising a bulge and a disc, both components are calculated separately, reflecting the different formation channels.
The resulting SEDs are saved in an HDF5 file, containing non-attenuated and dust-attenuated rest-frame absolute and observer-frame apparent magnitudes, for each filter and galaxy component.

For the generative mode of \prospect, the \citet{charlot2000} parameters must be given to \prospect.
In this work we have focused on two of the dust models presented in \citetalias{lagos2019}: CF00, which adopts the default parameters of \citet{charlot2000}, and T20-RR14\footnote{Called EAGLE-$\tau$ RR14 in \citetalias{lagos2019}.}, which uses the best-fit dust fraction-to-gas metallicity ratio from \citet{remy-ruyer2014} to calculate $\Sigma_\mathrm{dust}$, and then apply the  \citet{charlot2000} parameters $\Sigma_\mathrm{dust}$-dependency found in \citet{trayford2020} as detailed below.

In the T20-RR14 model, gas metallicities are used to calculate the dust-to-metal ratio, $f_\mathrm{dust}$, of each galaxy component, following the best-fit to the $\log_{10}(f_\mathrm{dust})$-$\log_{10}{Z}$ found by \citet{remy-ruyer2014}.
The dust mass derived from $f_\mathrm{dust}$ is then used to calculate the dust surface densities, $\Sigma_\mathrm{dust}$, for both bulge and disc.
For discs this is calculated as 
\begin{equation*}
    \Sigma_\mathrm{dust,disk}=M_\mathrm{dust,disc}/(2\pi r_\mathrm{50,disc}l_{50}),
\end{equation*}
where $M_\mathrm{dust,disc}$ is the metal mass in the disc, $r_\mathrm{50,disc}$ the half-mass radius of the gas in the disc (representing the major axis), and $l_{50}=r_\mathrm{50,disc}(\cos(i)(1-1/7.3)+1/7.3$ is the projected minor axis, with $i$ being the inclination of the galaxy. 
For bulges this is calculated as
\begin{equation*}
    \Sigma_\mathrm{dust,bulge}=M_\mathrm{dust,bulge}/(2\pi r_\mathrm{50,bulge}^2),
\end{equation*}
where $M_\mathrm{dust,bulge}$ is the metal mass in the bulge, $r_\mathrm{50,bulge}$ the half-mass radius of the gas in the bulge (assuming bulges to be spherical), and the $7.3$ factor comes from the relation between scale heights and lengths found in local disc galaxies \citep{kregel2002}.

The calculated optical depths for both discs and bulges account for the dust distribution in the ISM and in birth clouds, following the \citet{charlot2000} model, also used in \prospect, where:
\begin{align*}
    \tau_\mathrm{ISM}&=\hat{\tau}_\mathrm{ISM}(\lambda/5500\ A)^{\eta_\mathrm{ISM}}, \\
    \tau_\mathrm{BC}&=\tau_\mathrm{ISM}+\hat{\tau}_\mathrm{BC}(\lambda/5500\ A)^{\eta_\mathrm{BC}},
\end{align*}
where $X_\mathrm{ISM}$ are the quantities for the diffuse ISM screen, $X_\mathrm{BC}$ the quantities for the birth clouds, and $\hat{\tau}$ and $\eta$ denote the \citet{charlot2000} parameters for the corresponding screen.
The diffuse component is calculated following the $\hat{\tau}_\mathrm{ISM}$-$\Sigma_\mathrm{dust}$ and $\eta_\mathrm{ISM}$-$\Sigma_\mathrm{dust}$ relations found by \citet{trayford2020} in the radiative transfer post-processing of the EAGLE simulation, using the median and $1\sigma$ scatter of the relations as inputs for a Gaussian distribution from which $\hat{\tau}_\mathrm{ISM}$ and $\eta_\mathrm{ISM}$ are drawn.
For the birth clouds, the \citet{lacey2016} model is used, calculated as
\begin{equation*}
    \hat{\tau}_\mathrm{BC}=\hat{\tau}_\mathrm{ISM,0}\frac{f_\mathrm{dust}Z_\mathrm{gas}\Sigma_\mathrm{gas,cloud}}{f_\mathrm{dust,MW}Z_\odot\Sigma_\mathrm{MW,cloud}},
\end{equation*}
with $\hat{\tau}_\mathrm{BC,0}=1$, $f_\mathrm{dust,MW=0.33}$ and $\Sigma_\mathrm{MW,cloud=85}$ $M_\odot\mathrm{pc}^{-2}$, so that for typical spiral galaxies $\hat{\tau}_\mathrm{BC}\approx\hat{\tau}_\mathrm{BC,0}$.
The birth cloud gas surface density is defined as the maximum between $\Sigma_\mathrm{MW,cloud}$ and the gas surface density of either disc or bulge, $\Sigma_\mathrm{gas,cloud}=\mathrm{max}(\Sigma_\mathrm{MW,cloud},\Sigma_\mathrm{gas})$.
This is to account for galaxies with high ISM pressures, where $\Sigma_\mathrm{gas,cloud}\approx\Sigma_\mathrm{gas}$ \citep{krumholz2009}.

We refer the reader to \S2.1 and \S3.1 of \citetalias{lagos2019} for a more detailed description and discussion of these models.
Throughout the rest of this work we will refer to the LCs and galaxies made using the CF00 dust and extinction model as \shark\cf, and for those using the T20-RR14 models as \shark\trr.


\subsection{Building synthetic light-cones}\label{sec:lc_making}

All synthetic LCs used throughout this work have been created to match the footprint and magnitude selection of the equatorial GAMA fields.
To produce these LCs we have used  the publicly available code \sting\footnote{\url{https://github.com/obreschkow/stingray}} (Obreschkow et al. in prep.), an updated and extended version of the LC builder code used by \citet{obreschkow2009}.

\sting\ tiles the survey volume with a Cartesian grid of cubic simulation boxes in comoving coordinates.
The tiles are then populated with galaxies from the SAM snapshots that best match the lookback-time corresponding to the respective distance to the observer.
To avoid spurious coherent structures that may appear when the same galaxy is seen in multiple tiles, each tile uses a random symmetry operation (translation, rotation, inversion) -- a technique that exploits the periodicity of the simulation volume, as first described by \citet{blaizot2005}.
Our light cones extend to a distance equivalent to $z=0.6$, the maximum redshift used for the construction of the \gc.
This redshift limit corresponds to the last 30 snapshots of the L210N1536 simulation.

\begin{figure}
    \centering
    \includegraphics[width=\linewidth]{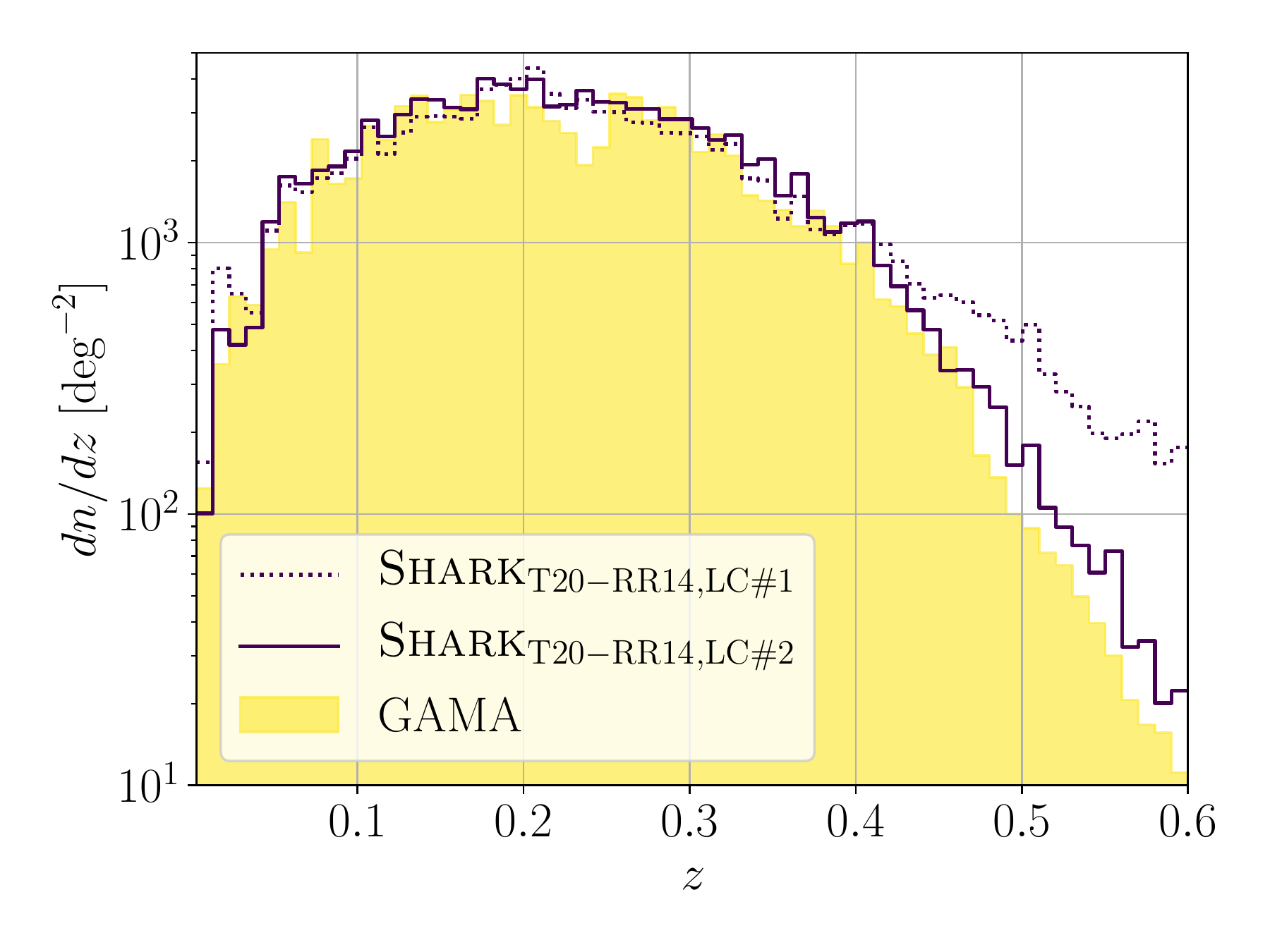}
    \caption{Redshift distribution of galaxies of LC sets \#1 and \#2 with GAMA. \shark\cf\ is not shown, as the distribution are nearly identical to those form \shark\trr. GAMA is shown with the solid histogram in yellow, \shark\trr\ the purple line, LC set \#1 with dotted lines and LC set \#2 with solid lines.}
    \label{fig:mock_nz}
\end{figure}

To avoid unnecessary use of resources, \sting\ allows the user to set thresholds on any set of galaxy properties from the simulation.
In addition, it also estimates a crude optical magnitude (simply called \texttt{mag} on the output files), which can also be used for a rough pre-selection before running \prospect\ for the final selection.
For this work we have selected galaxies with $M_\star>10^7$ $M_\odot$ and $\texttt{mag}<23.8$\footnote{Note that this meant changing in the \texttt{module\_user\_selections\_shark} file of \sting\ the default value of \texttt{dmag}, a value used to account for the scatter between \texttt{mag} and SED-produced magnitudes when using \texttt{mag} as as selection criteria, from 2.0 to 4.0, and of \texttt{selected} in the \texttt{sam\_selection} function for \texttt{case = 'gama'} from \texttt{sam\%mstars\_disk>1e8} to \texttt{(sam\%mstars\_disk+sam\%mstars\_bulge)>1e7}.}.
The choice of 23.8 was driven by comparing the \texttt{mag} values produced by \sting\ with the $r$-band values produced by \viper, which showed that any lower value at this stage would introduce magnitude incompleteness when making the final magnitude selection of $r<19.8$ from the \viper\ magnitudes.

\sting\ also computes the inclination for each galaxy relative to the observer, using the sub-halo angular momentum vector (as defined by \textsc{VELOCIRaptor}), under the assumption that the angular momentum vector of the galaxies points in the same direction.
This procedure is used for both \texttt{type 0} and \texttt{type 1} galaxies.
For \texttt{type 2}, satellites for which their sub-halo has been lost to \textsc{VELOCIRaptor}, usually due to becoming too small to be robustly identified \citep[see][for further information]{poulton2018}, their inclinations are randomly chosen.

To compare centrals and satellites in GAMA and our synthetic LCs, we follow the definition from the simulation, calling \texttt{type 0} galaxies centrals and merging both \texttt{type 1} (satellite) and \texttt{type 2} (orphan) galaxies into our satellite classification.
To emulate this classification, for the results from the \citetalias{robotham2011} group finder we assign all isolated and group centrals (\texttt{RankIterCen=0}) to our central classification, and all remaining galaxies (\texttt{RankIterCen>0}) as satellites.


\subsubsection{Synthetic LC set \#1: Direct comparison with simulations}\label{sec:LCv1}

By applying the same selection criteria as in GAMA ($r<19.8$) we produce the first set of synthetic LCs that we will use in this work, which we refer to as LC set \#1.
The redshift distribution for \shark\trr\ from this set is shown in Figure \ref{fig:mock_nz}, while \shark\cf\ is not shown as it closely resembles that from \shark\trr.
It is clear that, while producing a very good match at $z<0.4$, \shark\ over-estimates the number of galaxies at the high-redshift end.
This tension is not surprising as \shark\ slightly overestimates both the number density of massive galaxies and the cosmic star formation density at $z\sim0.5$ (both by $\sim0.3$ dex, see figures 2 and 5 of \citetalias{lagos2018}).
This is particularly relevant for the comparison at $z\ge0.4$, as at this redshift range the $r$ filter centre lies at $\sim410$ nm and the increased SFR leads to more galaxies reaching $r=19.8$ mag. 
We remind the reader that only galaxies brighter than $L^\ast$ being detected at $z\ge0.4$.

\begin{figure}
    \centering
    \includegraphics[width=\linewidth]{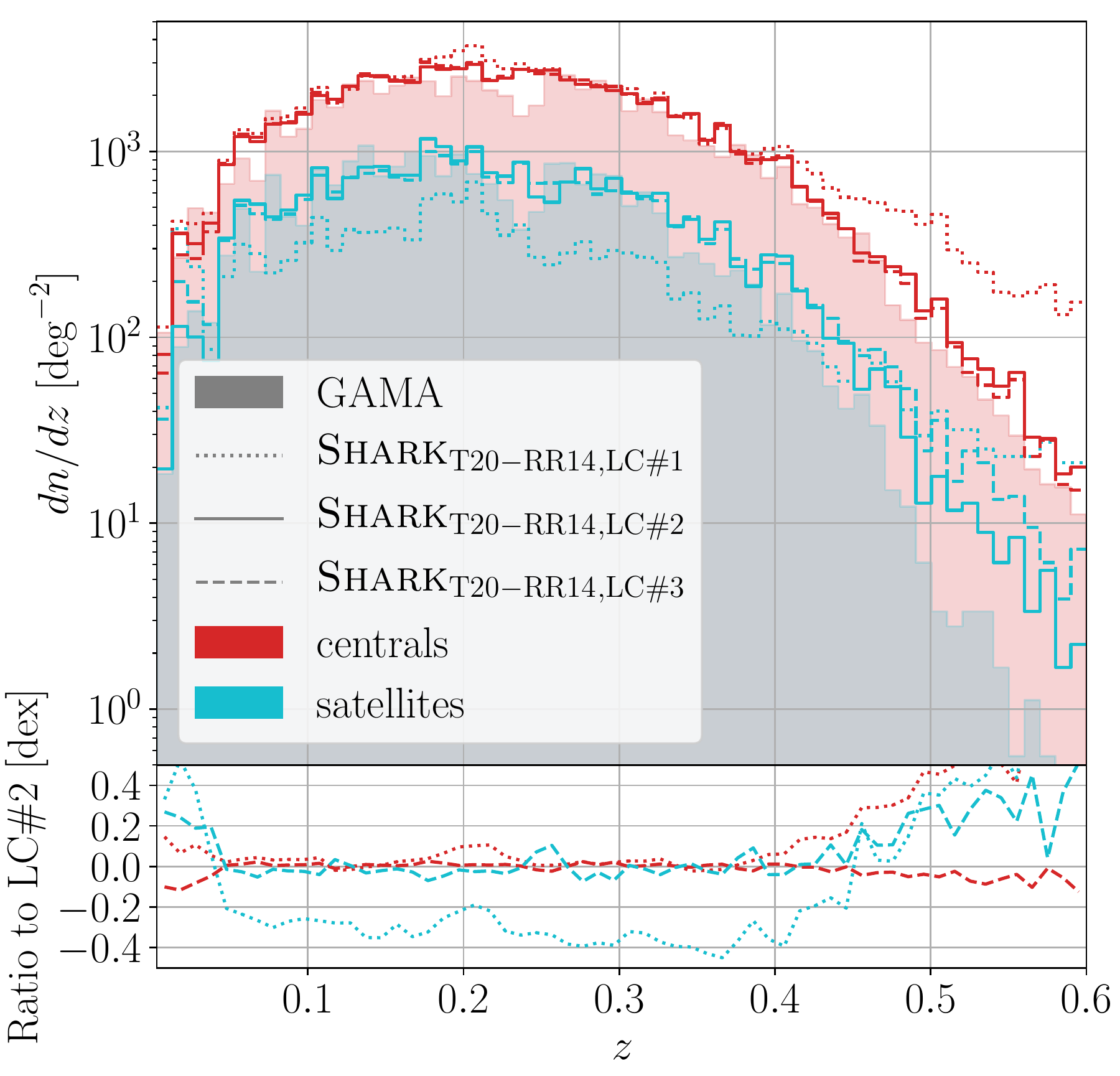}
    \caption{Redshift distribution of galaxies, separating central/satellite populations, of \shark\trr\ from LC sets \#1, \#2 and \#3 with GAMA. \shark\cf\ is not shown, as the distribution are nearly identical to those form \shark\trr. GAMA is shown as solid histograms, LC set \#1 with dotted lines, set \#2 solid lines and set \#3 with dashed lines. Blue histograms are for satellites and red for centrals. The top panel shows the distributions, and the bottom panel the logarithm of the ratio between sets \#1-\#3 and \#2.}
    \label{fig:mock_nz_censat}
\end{figure}

In the redshift range where the total distribution of our synthetic LCs are well matched to GAMA ($z\lesssim0.4$), \shark\ under-predicts the number of satellites by $\sim0.3$ dex, as can be seen in Figure \ref{fig:mock_nz_censat}.
At redshift above $\sim0.4$, the density of both populations in LC set \#1 surpasses the expected number from GAMA, with \shark\ grossly over-estimating the central (satellite) population by redshift $\sim0.55$, by a factor of $\sim10$ ($\sim30$).

\begin{figure*}
    \includegraphics[width=\textwidth]{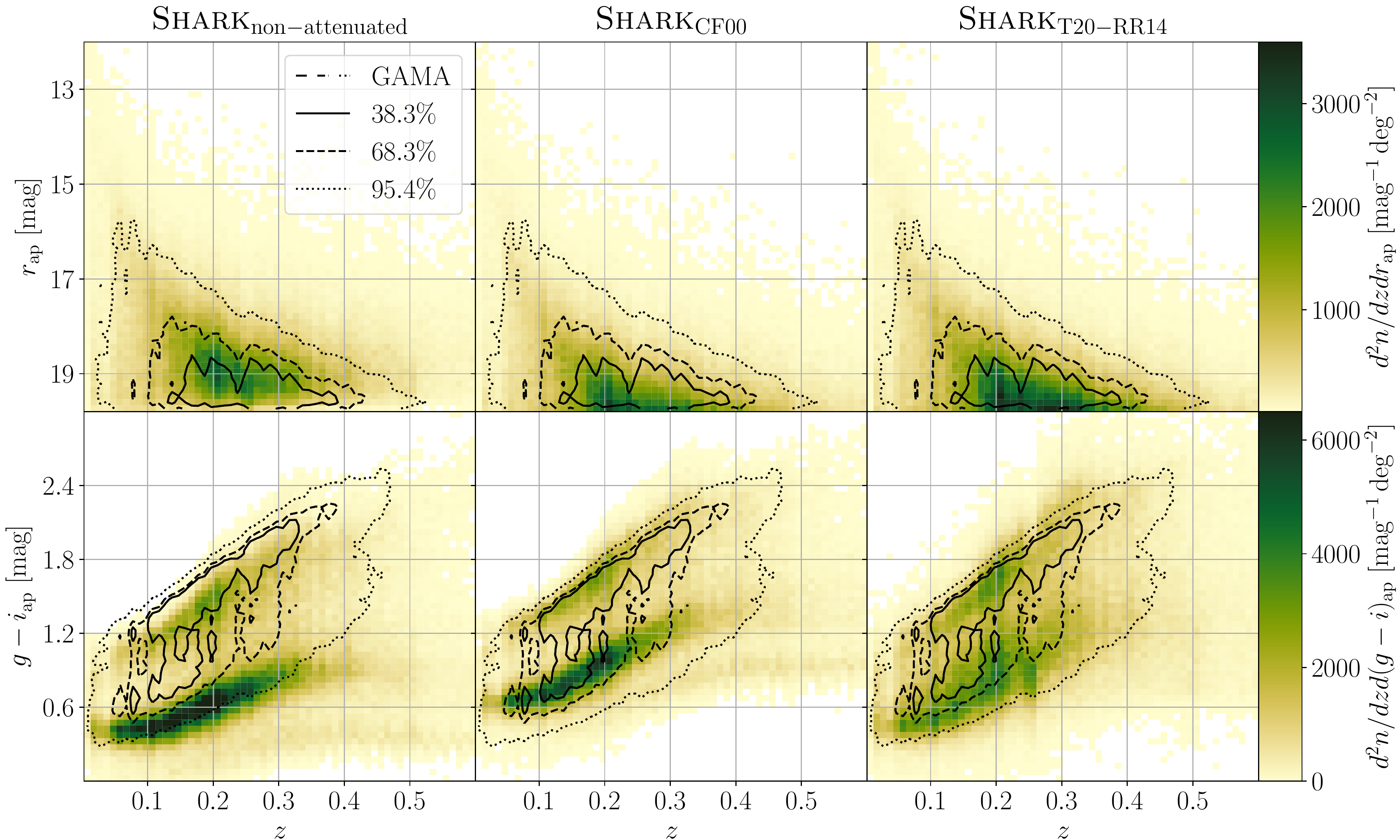}
    \caption{Magnitude (\rap) and colour (\giap) distributions from the LC set \#1 compared to GAMA. The left column shows the distributions for \shark\ non-attenuated photometry, the middle column for \shark\cf\, and the right column for \shark\trr. The galaxies shown on the left panel have been GAMA-selected using their \shark\trr\ \rap\ magnitudes.}
    \label{fig:mock_prop}
\end{figure*}

Figure \ref{fig:mock_prop} shows the \rap\ and \giap\ distributions as a function of redshift for \shark\ with no dust attenuation, \shark\cf\ and \shark\trr\.
The differences between the attenuation models in \shark\cf\ and \shark\trr\ are clear when compared with the non-attenuated distributions, with \shark\cf\ showing the simple magnitude/colour shift expected from using the same \citet{charlot2000} parameters for all galaxies, whereas \shark\trr\ produces a more complex change that ends up closer to the observed distribution.
Noticeable is that in the non-attenuated colour distribution there is a small subset of the blue population that branches off by $z\sim0.25$, generating two parallel blue populations, with a separation of $\sim0.5$ mag.
While \shark\cf\ produces a fairly good match to observations, the blue and red populations are more distinct than in GAMA, and the branching blue subset remains present.
In contrast, \shark\trr\ produces more green-valley galaxies, at the cost of a blue population that is slightly too blue ($\sim0.2$ mag).
The reason for the bluer ``blue cloud'' in \shark\trr\ compared to \shark\cf\ is because the former cares about dust surface density.
\citet{lagos2018} showed that \shark\ tends to slightly underestimate the gas metallicities of galaxies with $M_\star\lesssim10^{10}M_\odot$, which results in dust masses that are slightly too low.
The implication is therefore that the optical depth of these galaxies is smaller than we would expect for more metal-rich galaxies, yielding bluer colours.
Also note that the branching blue subset has almost completely disappeared on \shark\trr.
Visible on both models on their \rap\ distributions is the presence of some extremely bright galaxies, up to $\sim4$ mag brighter than the $95.4\%$ contours from GAMA.

{Detailed analysis of the galaxies in the blue population branch shows that they are a product of artefacts in the merger tree, due to halos whose branch gets erroneously assigned to a nearby halo at a cosmic time just prior to their appearance on the LC. From the perspective of \shark\, these halos have just `popped up', and therefore are assumed to be a new structure with pristine gas. Because these halos can be quite massive ($>10^{11}M_\odot\mathrm{h}^{-1}$) they undergo a large burst of star formation, explaining their extremely blue colour (i.e. large SFRs and low metallicity). This issue also becomes apparent with the tree parameter measurement defined by \citet{obreschkow2020}, where an excess of halos dominated by smooth accretion can be seen in their Figure~8. We choose not to remove these galaxies, because only a small amount of halos are affected \citep[$0.7\%$ for $M_\mathrm{halo}>10^{11}M_\odot\mathrm{h}^{-1}$;][]{obreschkow2020}, and because that branch merges with the ``normal'' blue cloud in \shark\trr.}


\subsubsection{Synthetic LC set \#2: Abundance matching and accounting for observational errors}\label{sec:LCv2}

\begin{figure*}
    \centering
    \includegraphics[width=\linewidth]{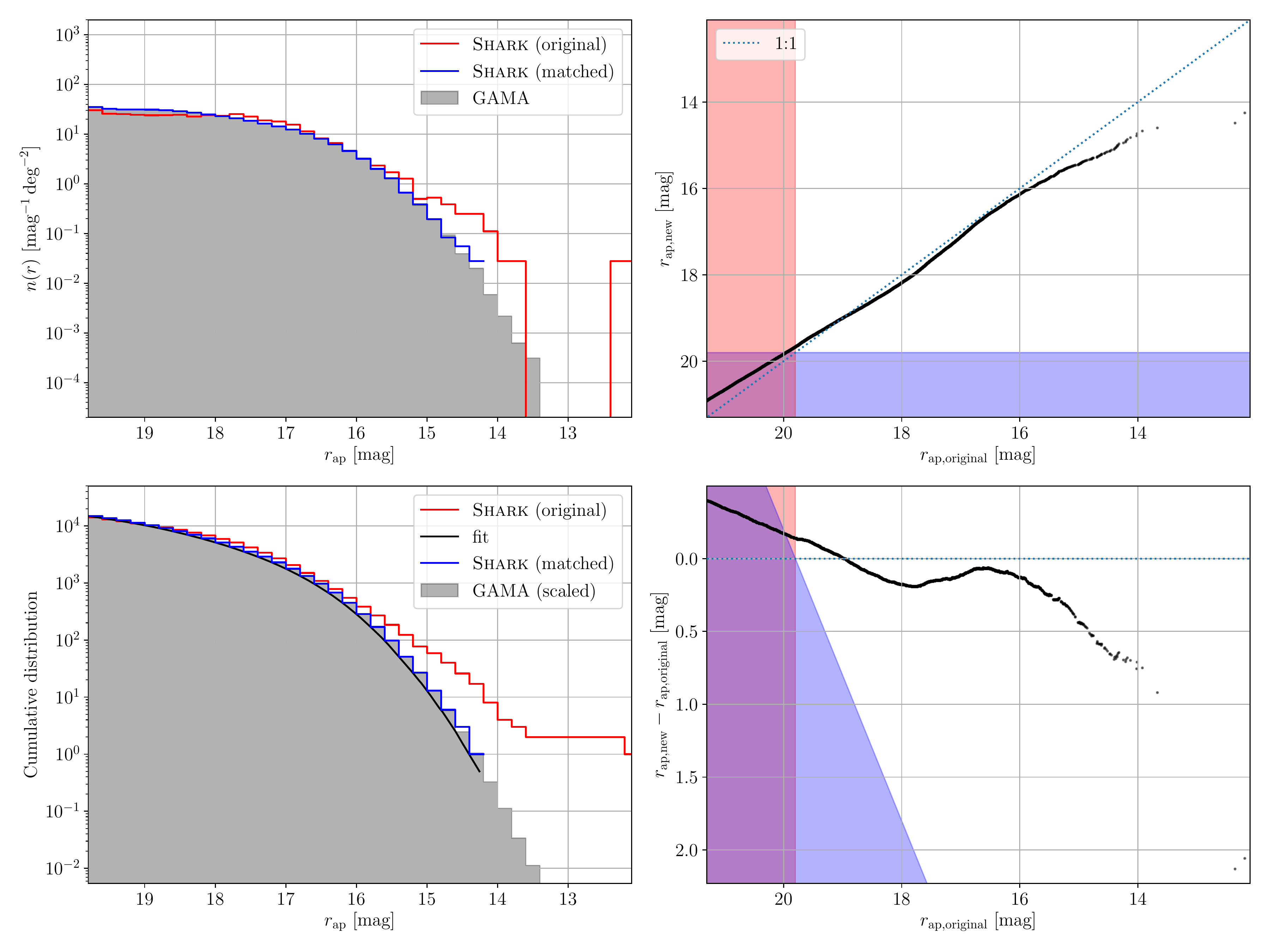}
    \caption{Diagnostic plot for the abundance matching procedure in the redshift bin $0.0703<z<0.1078$. Clockwise from the top left: The \rap\ magnitude distribution of GAMA, T20-RR14 photometry prior abundance matching and the new T20-RR14 photometry; the abundance-matched $r_\mathrm{ap,new}$ magnitude as a function of the original $r_\mathrm{ap,original}$; the difference between abundance-matched and original magnitudes as a function of the original; and the cumulative distributions for GAMA and both T20-RR14 photometries, together with the spline fit (black line) to the GAMA distribution used to draw the abundance-matched values. On the left panels, the GAMA random catalogue is shown by the solid grey histogram, \shark\ intrinsic distribution by the red line, and the abundance-matched version of \shark\ by the blue line. On the right panels, the red-shaded region represents the GAMA limit of $r<19.8$ applied to the original magnitudes,while the blue-shaded region is the same but for the abundance-matched magnitudes.}
    \label{fig:abmatch}
\end{figure*}

As having synthetic LCs that reproduce the average galaxy density at any given redshift range is critical for many purposes, such as the calibration of group-finder algorithms, we decide to further fine-tune our LCs by re-building the selection band photometry using an abundance matching method.
For this, we have used the publicly-available random galaxy catalogue for the G15 field\footnote{\url{http://www.gama-survey.org/dr3/schema/dmu.php?id=19}} produced with the procedure described in \citet{farrow2015}.
The benefits of using this catalogue are that this sample has been constructed to remove the large-scale structure variations observed in the survey, and the galaxy replication provides for an ample number of galaxies to use a fine binning in redshift, even near the redshift limit of $z=0.6$, where the need for the abundance matching becomes critical to solve the tension between \shark\ and GAMA.

To perform the abundance matching we divide this random sample into 16 redshift bins, each with a width of 0.0375.
For each redshift bin we calculated the cumulative distribution of galaxies as a function of their $r$ apparent magnitudes, in 250 bins of 0.0472 mag.
We fit a cubic spline to the resulting distributions\footnote{Using the \texttt{PchipInterpolator} function from the \textsc{SciPy} Python package, with extrapolation, enabled (\texttt{extrapolate=True})}, fitting magnitude as a function of the number of galaxies, to avoid the integration step necessary if fitting the number of galaxies as a function of magnitude.
This choice of binning the highest resolution possible with well-behaved interpolations with the method used.
Decreasing the number of redshift bins negatively affected the performance of the abundance matching at high redshift ($z\gtrsim0.4$), and either decreasing the width or making it variable for magnitudes both produced fits that would flip on the extrapolation regime.

\begin{figure}
    \centering
    \includegraphics[width=\linewidth]{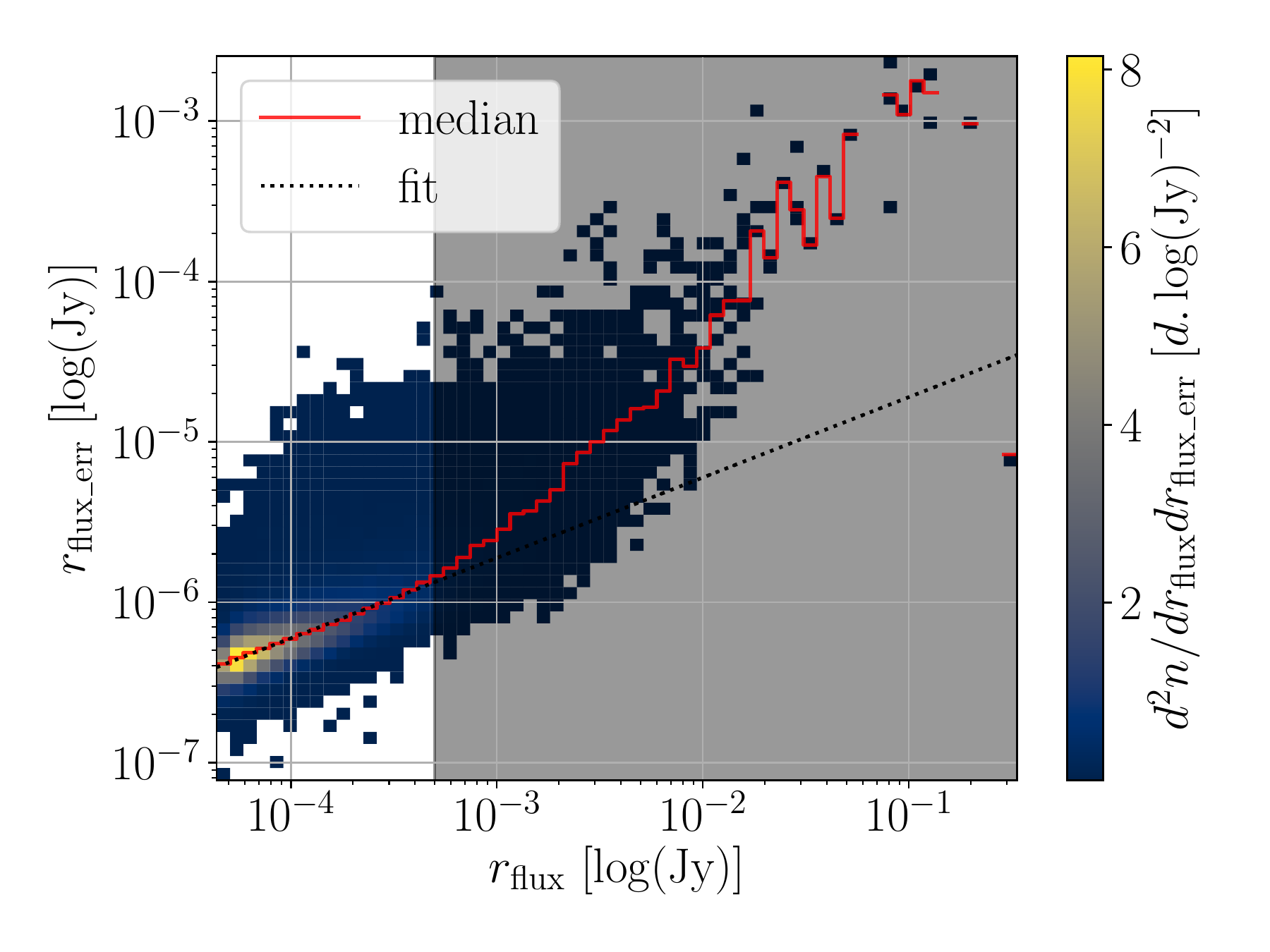}
    \caption{Flux error distribution for the $r$ band in GAMA displayed as SNR as a function of flux. The red solid line shows the running median of the distribution. The dotted black line the cubic spline fit the running median that we use to model the photometric errors in our synthetic LC sets \#2 and \#3.}
    \label{fig:flux_err}
\end{figure}

Using these fits we calculate new magnitudes for all galaxies with $r<21.3$ in our synthetic LCs, to ensure both completeness down to $r=19.8$ when adding errors to the magnitudes and that enough galaxies are available if any of the LCs is under-dense compared to the GAMA random catalogue.
Figure \ref{fig:abmatch} presents an example of the diagnostic plots we produced for this process, showing the abundance matching results for the $0.0703<z<0.1078$ redshift bin for \shark\trr, where it is clear that the brightest galaxies seen in Figure \ref{fig:mock_prop} need to be corrected down by $\sim2$ mag to be brought in agreement with GAMA.

To keep consistency between the different galaxy properties while avoiding implied and/or explicit changes in the distribution of stellar populations of our synthetic galaxies, we adjust the magnitudes in all other filters by the difference between the \viper\ magnitudes in the selection filter and the abundance-matched values (i.e., leaving colours unchanged).

Furthermore, we scale the stellar masses of all galaxies by the factor implied by these magnitudes changes:
\begin{equation*}
    \log_{10}(M_{\star,\mathrm{match}})=\log_{10}(M_{\star,\mathrm{ref}})-(m_\mathrm{match}-m_\mathrm{ref})/2.5,
\end{equation*}
\noindent where $m$ represents the $r$ magnitude for the synthetic LC, quantities with the $_\mathrm{ref}$ subscript are from the simulation, and those with the $_\mathrm{match}$ subscript are from the abundance matching.
For the example shown in Figure \ref{fig:abmatch}, this implies a change in stellar mass for most galaxies, with $|r_\mathrm{ap,new}-r_\mathrm{ap,original}|<0.5$, of at most 0.2 dex.
The two most massive galaxies undergo a more significant change, with the stellar masses reduced by $\sim0.8$ dex.

Measurement uncertainties, if not affected by biases, will broaden and mix the observed distributions of galaxy properties, and to replicate this effect we add empirically-motivated errors to our synthetic LCs.
We use the reported errors for the measured flux and stellar masses for GAMA to model the errors in our synthetic LCs.

\begin{figure}
    \centering
    \includegraphics[width=\linewidth]{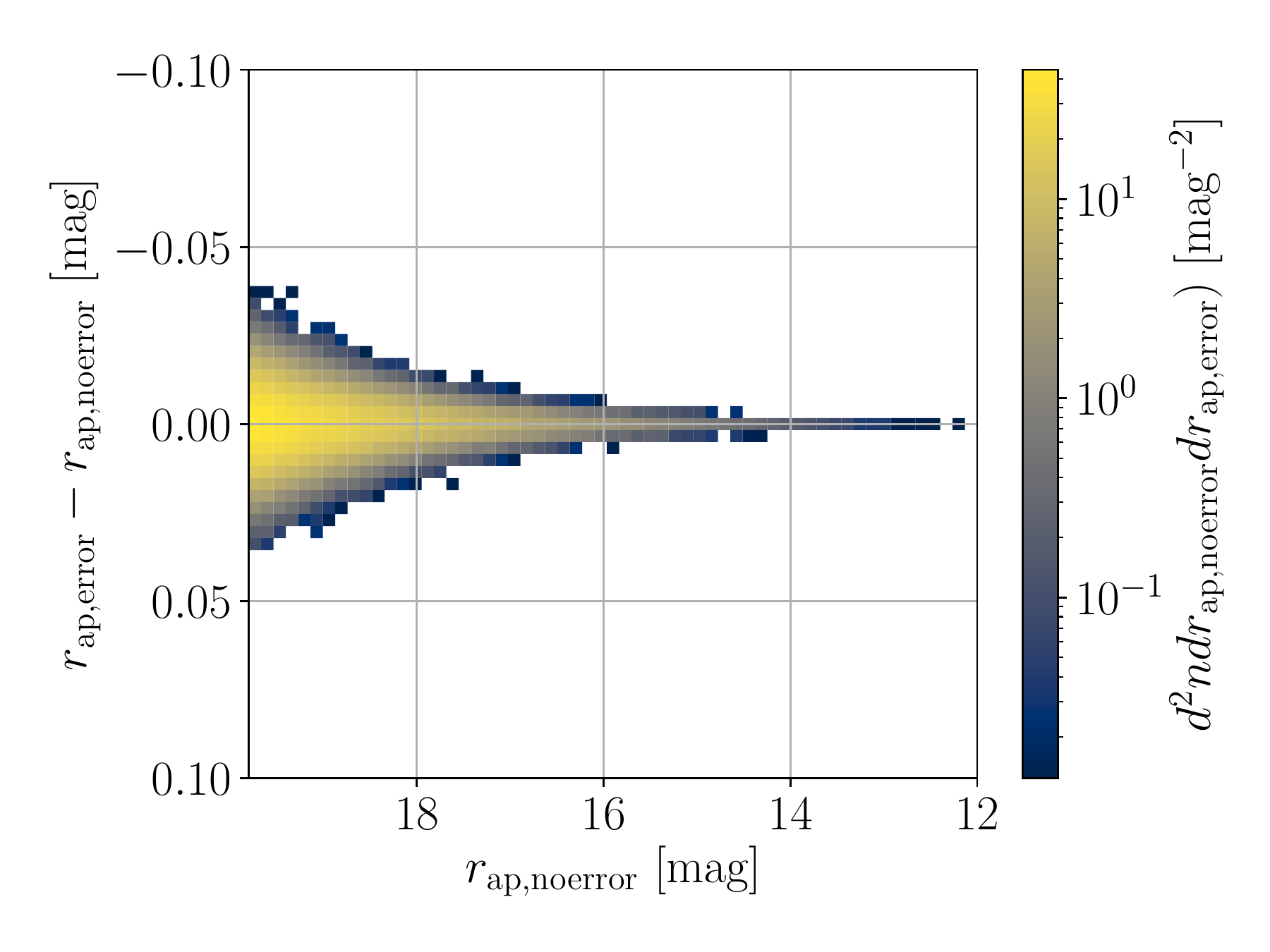}
    \caption{Distribution of the magnitude errors applied to the $r$ band of LC set \#2, as a function of the $r$ band magnitude prior application of the errors. The shaded region was not used for the modelling of the error.}
    \label{fig:mag_err}
\end{figure}

In GAMA the errors in stellar masses are consistent with a constant uncertainty of $\approx0.11$ dex, so we perturb every galaxy stellar mass in our synthetic LCs by a factor of $10^{\sigma_{M_\star}}$, with $\sigma_{M_\star}$ being a value drawn from a Normal distribution with  $\mu=0$ and $\sigma=0.11$.
The uncertainties for the photometry exhibit a more complex behaviour, as seen in Figure \ref{fig:flux_err}.
{While from a simple argument one would expect for the logarithm of the noise to scale linearly with the logarithm of the flux, it is clear that for brighter galaxies in the \textsc{KidsVikingGAMA} catalogue the noise deviates from this expectation. This is driven by the way noise measurement is computed by \textsc{ProFound}. Normally, the noise depends on the galaxy flux. However, for the large segments occupied by nearby bright galaxies, the dominant component of the noise is instead the estimated sky flux error. For this reason we excluded galaxies above $0.5$ mJy for the modelling of the photometric error for the application to \shark. For the latter, we compute the running median and fit a line in log-log space to this median. We then use this linear relation to calculate the model flux error.}
We then perturb the fluxes in each band in \shark\ by a flux error drawn from a Normal distribution with $\mu=$ and $\sigma=\sigma_\mathrm{fit}(f)$, where $\sigma_\mathrm{fit}$ is the modelled error and $f$ is the flux of each galaxy.
For filters other than $r$ we use the same fit, which we extrapolate to provide the error fit for flux values below the $r$ band flux limit.
The difference between magnitudes with and without this model errors are shown in Figure \ref{fig:mag_err}.

We do not apply errors to the redshifts to any of our synthetic LCs, as relative to the scales and bins used in this work the uncertainties associated with spectroscopic redshifts are negligible. We refer to these LCs as LC set \#2.

While by construction our abundance matching means that our synthetic LCs are in good agreement with the distribution from GAMA for the entire galaxy population, as displayed in Figure \ref{fig:mock_nz}, Figure \ref{fig:mock_nz_censat} shows that tension remains in the number distribution of satellite galaxies.
This naive comparison of the properties of centrals and satellites between our observed and synthetic samples ignores the differences between how this classification is defined in both cases, which is especially relevant for SAMs like \shark\ where galaxies evolve following different physical prescriptions if they are classified as a central or satellite.

\begin{figure*}
    \includegraphics[width=\textwidth]{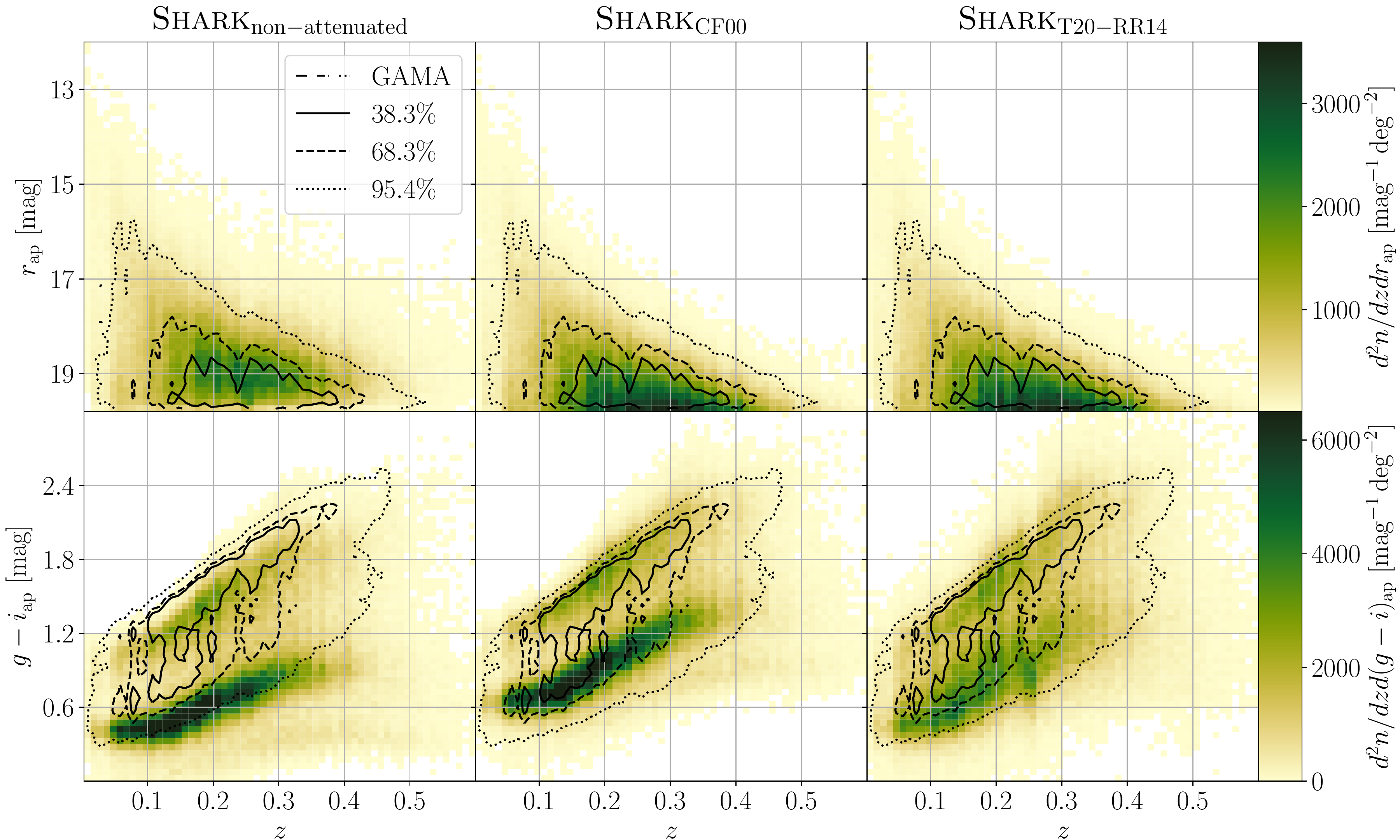}
    \caption{Magnitude (\rap) and colour (\giap) distributions from the LC set \#2 compared to GAMA. The magnitudes have been abundance-matched, and empirically-motivated errors have been applied. The left column shows the distributions for \shark\ non-attenuated  photometry, the middle column for \shark\cf\, and the right column for \shark\trr. The galaxies shown on the left panel have been GAMA-selected using their \shark\trr\ \rap\ magnitudes.}
    \label{fig:mock_abmatch_prop}
\end{figure*}

The resulting \rap\ and \giap\ distributions as a function of redshift are shown in Figure \ref{fig:mock_abmatch_prop}. 
The addition of errors has slightly broadened both blue and red populations in \shark\cf, and while the abundance matching has reduced the number of galaxies at high redshift, the branching of the blue population remains visible.
For \shark\trr\ there is no discernible difference between the colour distributions besides the decreased number of high-redshift galaxies.
The effect of the abundance matching is clearly seen in both models in the \rap\ though, with noticeably dimmer galaxies at the bright limit of the distributions, as expected from $\sim2$ mag reduction seen for the brightest galaxies on Figure \ref{fig:abmatch}.

For GAMA, the best definition is provided by the iterative ranking method defined in \citetalias{robotham2011}, where once the Friend-of-Friends (FoF) algorithm defines which galaxies belong to which group, an initial estimate of the centre of the group is made by calculating the centre-of-luminosity.
Then it iterates by removing the most distant galaxy and re-calculating the centre-of-luminosity, until two galaxies remain, of which the brightest one is defined as the group central.
The reliability of this method is reduced as the number of members in the group decreases, which means that in groups with a low number of members (either because of size or survey limits), diminishing the differences between centrals and satellites compared to both the ground truth and our synthetic LCs.

Furthermore, the group finder used by \citetalias{robotham2011} was trained on a simulation where halos, and hence galaxy groups, were defined using a 3D FoF \citep[\textsc{SubFind},][]{springel2001}, while in the simulation from which we created our synthetic LCs the halos were defined using a two-stage finder \citep[\textsc{VELOCIraptor},][]{elahi2018a}, which first produces a halo catalogue using a 3D FoF, then makes a second pass using a 6D FoF to separate kinematically distinct structures grouped by the 3D FoF.
This would have the effect of reducing the number of satellites/increasing the number of centrals, relative to a pure 3D FoF, which is consistent with what is shown in Figure \ref{fig:mock_nz_censat}.

To make a fair comparison between observations and simulations, we have used the \citetalias{robotham2011} group finder to generate the central/satellite classification for LC set \#2, after doing the abundance matching and addition of errors, with the same calibration as the one used for \gc\ (see Figure \ref{fig:flow_chart} for a schematic view of this process).
Appendix \ref{sec:GFcheck} show the most informative quality checks on this group finding.
This noticeably reduces the tension between observation and simulation, where now our synthetic LCs are well matched to GAMA up to $z\sim0.4$, as seen in Figure \ref{fig:mock_nz_censat}.
This shows that the exact definition of satellite/central in our observations is not one-to-one with those in simulations. This may not be a surprising statement, but in practice these definitions are used in the literature to assess how well simulations reproduce observations.
Our analysis shows that this comparison should be treated carefully.
Despite the success of \shark\trr, the match is not perfect.
Above $z\sim0.4$ the number of satellites is still slightly over-predicted, with both \shark\cf\ and \shark\trr\ having $\sim3$ times more satellites than GAMA.


\subsubsection{Synthetic LC set \#3: Confused \shark\ central-satellite classification}\label{sec:LCv3}

While running a group finder on synthetic LCs yields the best reproduction of observations, given the extra computational cost involved it is worthwhile testing faster alternatives.
At first order, it is reasonable to expect for the central/satellite classification errors to be driven by the sparse sampling that the detected galaxies provide of the total mass distribution of a halo. If halo/galaxy properties do not play a significant role on the likelihood of a galaxy being misclassified, a random re-assignment of a fraction of the centrals/satellites in a synthetic LC as satellites/centrals should reproduce the observed effects of misclassification.

To this end we created a third LC set, taking the LC set \#2 from the previous section and, instead of using the \citetalias{robotham2011} group finder central/satellite classification, we used the simulation classification and randomly re-assigned $15\%$ of the centrals/satellites as satellites/centrals.
This percentage was chosen as it provides a good match to the redshift distribution of both populations on GAMA, as seen in Figure \ref{fig:mock_nz_censat}.
A comparison between LC sets \#2 and \#3 in Figure \ref{fig:mock_nz_censat} shows that the satellite numbers produced by the random re-assignment are in good agreement with those from the \citetalias{robotham2011} group finder for $z\lesssim0.45$.
This is not a significant issue though, as the $0.45<z<0.6$ range only contains $\approx2\%$ of the galaxies (for GAMA and the abundance-matched LCs) while being $\sim50\%$ of the comoving volume, which makes this range of limited value.

\begin{figure*}
    \centering
    \includegraphics[width=\linewidth]{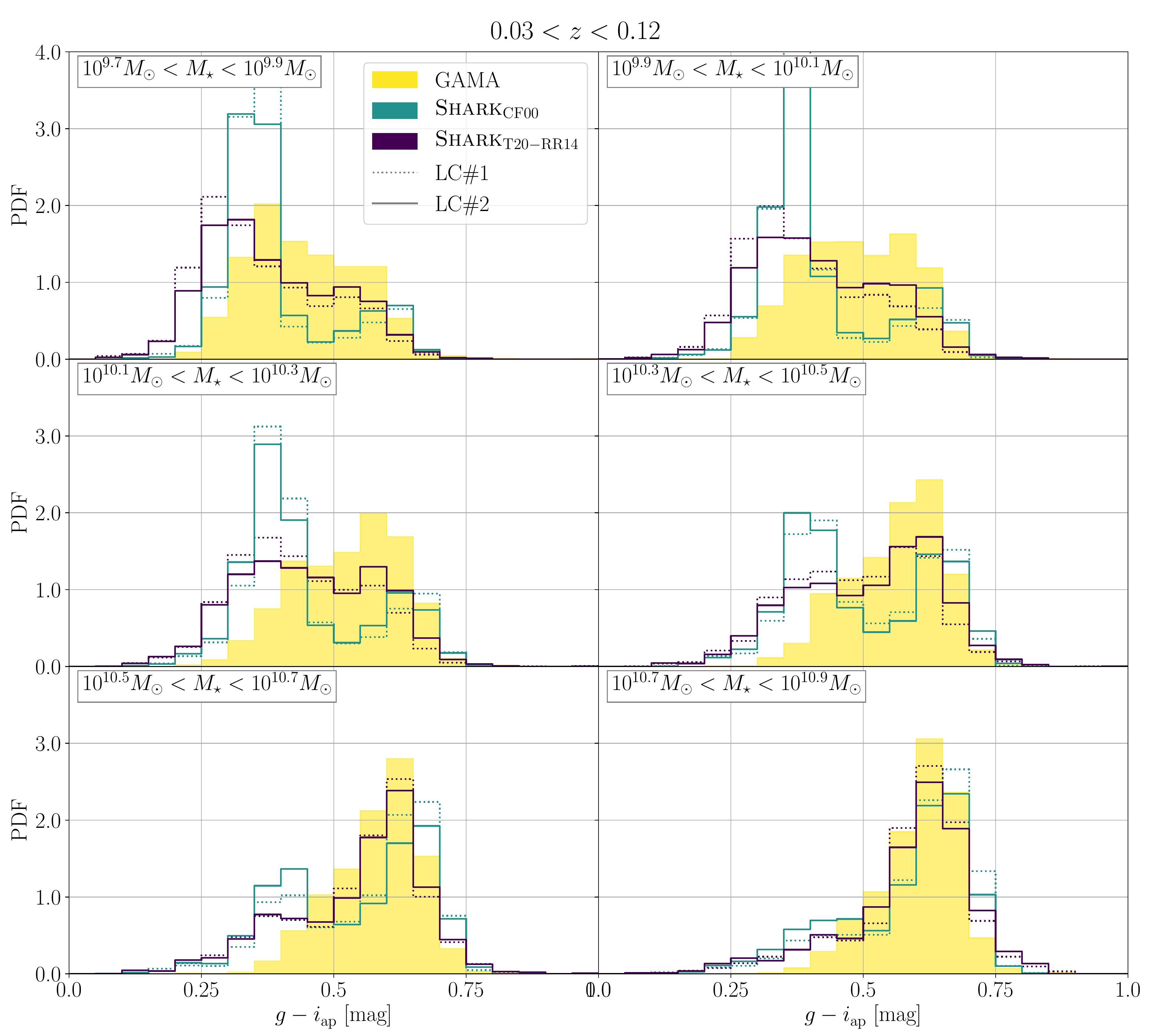}
    \caption{Apparent observer-frame $g-i$ colour distribution of galaxies with $0.03<z<0.12$ in LC sets \#1, \#2 and GAMA. The histograms are shaded/coloured as in Figure \ref{fig:mock_nz}, with \shark\cf\ now shown with the green lines. The stellar mass range of each bin is shown on the top left of each panel.}
    \label{fig:gi_z008_dist_all}
\end{figure*}


\section{Comparison of colour distributions in \shark\ and GAMA}\label{sec:colours}

To compare the colours of galaxies of our synthetic LCs to GAMA we use a similar selection to the one used in figure 6 by \citet{taylor2015}, where we choose galaxies with $0.03<z<0.12$ and divide them in stellar mass bins of $0.2$ dex.
While GAMA extends further in redshift, with the median being $z\sim0.2$, the $r<19.8$ limit means that GAMA is mass-complete only for galaxies with $M_\star\gtrsim M^\ast\sim10^{10.5}M_\odot$ by $z\sim 0.2$.
Only the low redshift range is therefore capable of capturing the galaxy populations on both sides of the characteristic mass $M^\ast$.
We choose stellar masses between $10^{9.7}M_\odot$ and $10^{10.9}M_\odot$, the lower end to ensure that our sample is complete given GAMA survey limits, and the upper to ensure the selection of at least 20 satellite galaxies from GAMA and both \shark\cf\ and \shark\trr.
For the colour comparisons in \S\ref{sec:full_pop} and \S\ref{sec:censat}, we have chosen to use observer-frame apparent magnitudes.
Though this means we are probing different parts of the SED of a galaxy as a function of redshift (though the shift at $z=0.12$ is not significant), observed colours are among the most direct observables, and thus the success or failure of our synthetic LCs in matching the distributions in GAMA is driven by the models adopted in our simulations and the assumptions made when doing abundance matching (only for LC sets \#2 and \#3).
In \S\ref{sec:passives} we switch to rest-frame absolute magnitudes, as to define a galaxy as red or passive based on colour is necessary to probe the same region of the SEDs for all galaxies.


\subsection{Full population distributions}\label{sec:full_pop}

As \citetalias{lagos2019} used rest-frame absolute magnitudes for their analysis, it is informative to first compare the global colour distributions of \shark\cf\ and \shark\trr\ before delving on the distributions for centrals and satellites.
First we will present the results using our synthetic LC set \#1 (\S\ref{sec:LCv1}), followed by those of synthetic LC set \#2 (\S\ref{sec:LCv2}).
Since the only change between LC sets \#2 and \#3 is the switch from the group finder classification of centrals and satellites to the random mixed one, we have not included the latter in this part of the analysis.

Figure \ref{fig:gi_z008_dist_all} shows the PDF of the $g-i$ colour index distributions of GAMA and LC sets \#1 and \#2.
At the high-mass end of our sample ($10^{10.7}M_\odot<M_\star10^{10.9}M_\odot$) both photometry sets are in very good agreement with GAMA, while below that stellar mass both models start to diverge.
\shark\cf\ shows a clear bimodal distribution for the rest of the stellar mass bins, which is not observed in GAMA, with the blue population being bluer than GAMA and the main peak being slightly redder.
We find that the transition of galaxies from blue-dominated to red-dominated in \shark\cf\ is at $10^{10.5}M_\odot<M_\star<10^{10.7}M_\odot$ range. This is $\sim 0.4$~dex higher stellar mass than the observed transition in GAMA.
Interestingly the main peak becomes consistent again with GAMA at the lowest mass bin, though the shape of the distributions remains in strong tension.

\shark\trr\ appears to more closely reproduce the shape of the colour distributions observed in GAMA. However, some areas of tension remain. 
The blue cloud peak at stellar masses $<10^{10}\,\rm M_{\odot}$ happens at a $g-i$ that is $\sim0.1$~mag bluer than observed in GAMA. We also find that the transition from blue- to red-dominated galaxy populations happens at a lower stellar mass than for \shark\cf, at $10^{10.3}M_\odot<M_\star<10^{10.5}M_\odot$. Although this is an improvement over \shark\cf, still is $0.2$~dex too high stellar mass than where the transition happens in GAMA.

\begin{figure}
    \centering
    \includegraphics[width=\linewidth]{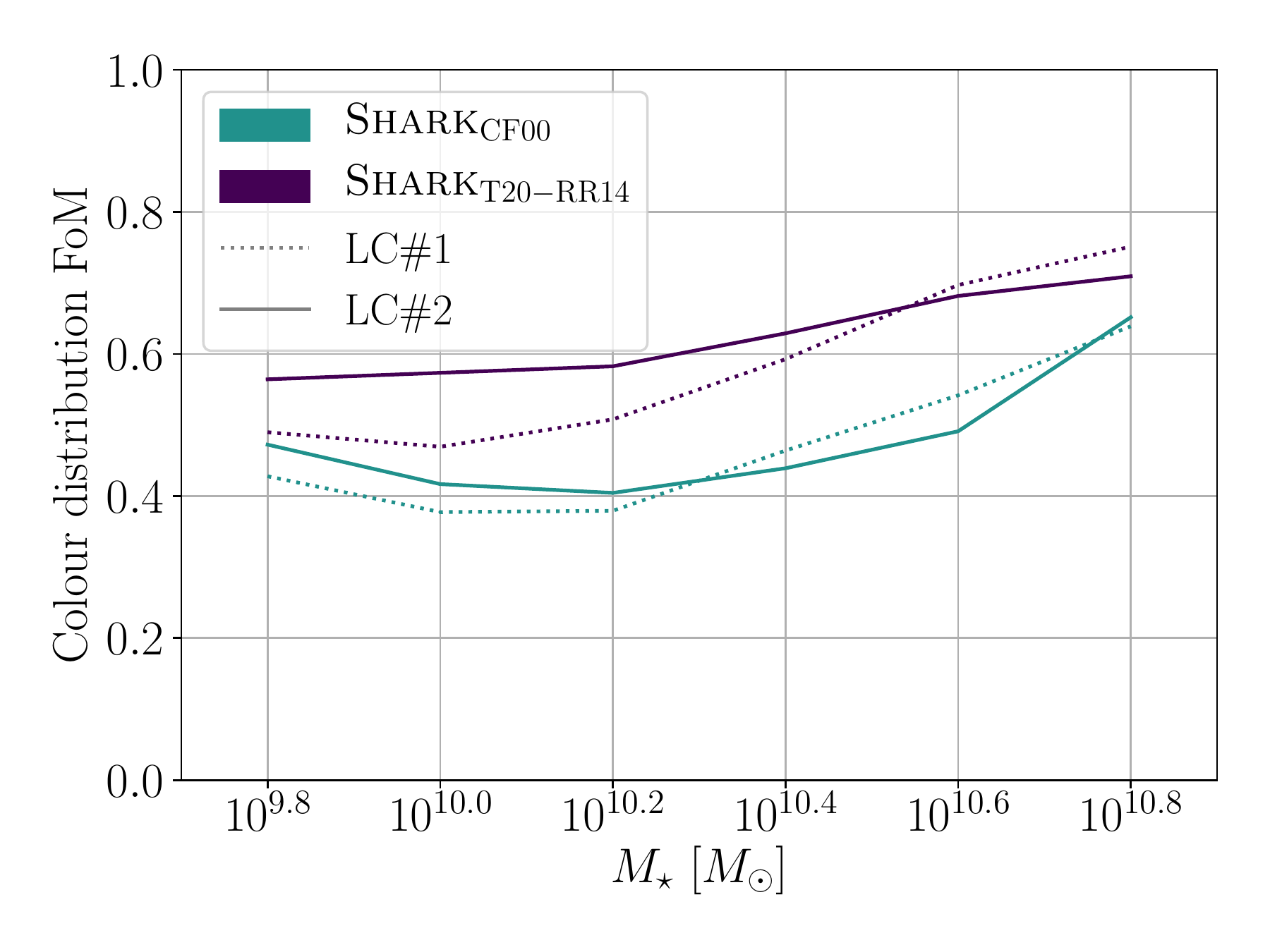}
    \caption{Colour distribution FoM of the colour distributions of both LC sets \#1 and \#2 for both \shark\cf\ and \shark\trr\. FoM defined as the division of the area of the intersection between the photometry and GAMA, divided by the area of the union of both distributions. Line colours as in Figure \ref{fig:mock_nz}.}
    \label{fig:gi_z008_match_all}
\end{figure}

The effect of performing the abundance matching and adding observational errors is also shown in Figure \ref{fig:gi_z008_dist_all}.
The results for \shark\cf\ are mostly unchanged across all stellar mass bins, with the main difference being that at the low-mass end (top row) there is a slightly higher peak for the blue population.
In contrast, while the issue of the bluer-than-GAMA peak for the blue population of \shark\trr\ at the low-mass end remains, there is an improvement by performing the abundance matching and addition of errors.
The transition between blue- to red-dominated distributions is closer to GAMA, becoming more apparent at $10^{10.1}M_\odot<M_\star<10^{10.3}M_\odot$. However, we still find that we need to go higher in stellar mass to $10^{10.3}M_\odot<M_\star<10^{10.5}M_\odot$ to see the galaxy population being clearly red-dominated.

To quantify the quality of the match between either \shark\cf\ or \shark\trr\ to GAMA we have calculated a Figure of Merit (FoM) by dividing the area of the intersection between the \shark\cf$_/$\trr\ and GAMA PDF areas by the union of said areas\footnote{The value of our FoM correlates to the $p$-value produced by a two-sample Kolmogorov-Smirnov test, but it provides a more meaningful measurement, as the imperfect nature of our models lead to $p$-values that are always too small for a significant comparison.}.
For this measurement, a value of 1 would represent a perfect match, while a value of 0 would represent completely disjointed distributions.
These values can be seen in Figure \ref{fig:gi_z008_match_all}, where we have plotted them as a function of stellar mass.
These results reinforce our analysis that \shark\trr\ is the superior model, but it is important to remark that it can only complement the qualitative analysis, as this FoM does not capture well the shape of the colour distribution.
After the abundance matching and addition of errors \shark\cf\ remains mostly the same, but the difference with \shark\trr\ from for masses below $10^{10.5}M_\odot$ has widened, reflecting the better fit it provides to GAMA.

This tells us that \shark\ galaxies are slightly too bright/massive for their colours, but they have SFHs and ZHs that more or less produce the right stellar populations in galaxies around M$^{\ast}$ and above.
Below M$^{\ast}$ the issue becomes the gas metallicities of low-mass galaxies being slightly too low compared to observations, leading to the bluer peak in both \shark\cf\ and \shark\trr, as already discussed above.


\subsection{Central and satellite population distributions}\label{sec:censat}

The analysis in the previous section shows that \shark\trr\ is the more successful model of the two, so for clarity in the figures on this section, we will not show the results of \shark\cf.
As already stated earlier, we are calling \texttt{type 0} galaxies centrals and grouping both \texttt{type 1} and \texttt{type 2} galaxies as satellites.
For the \citetalias{robotham2011} group finder of GAMA, we assign all \texttt{RankIterCen=0} galaxies as centrals, and the rest (\texttt{RankIterCen>0}) as satellites.

\begin{figure*}
    \centering
    \includegraphics[width=\linewidth]{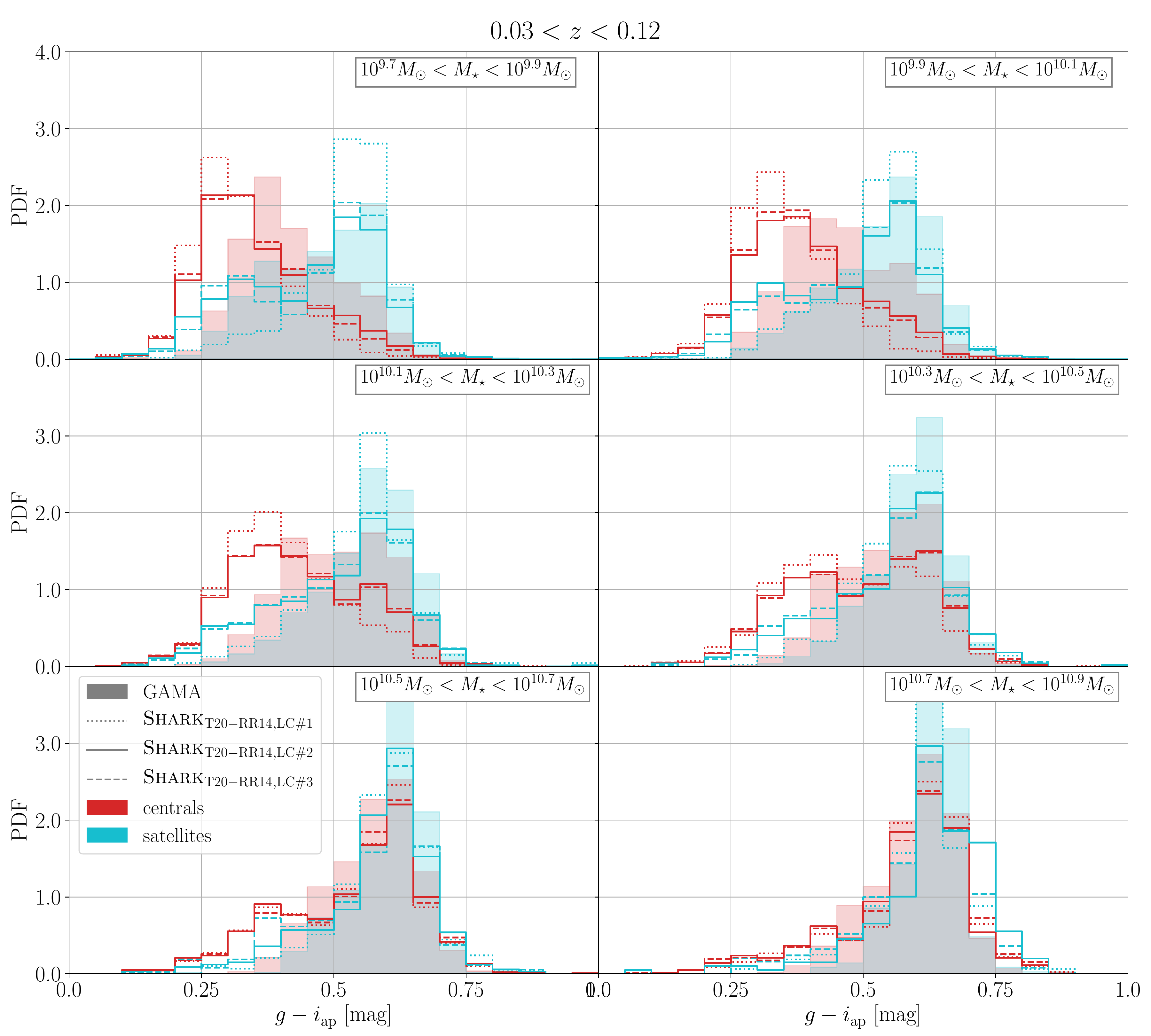}
    \caption{Apparent observer-frame $g-i$ colour distribution of galaxies with $z<0.12$ in LC sets \#1, \#2, \#3 and GAMA, divided into centrals and satellites. The histograms are shaded/coloured as in Figure \ref{fig:mock_nz_censat}. The stellar mass range of each bin is shown on the top left of each panel.}
    \label{fig:gi_z008_dist_censat}
\end{figure*}

Figure~\ref{fig:gi_z008_dist_censat} follows the same panel structure as Figure~\ref{fig:gi_z008_dist_all}, but with galaxies from GAMA and \shark\trr\ split between centrals and satellites.
Due to being the dominating type by number, the colour distributions of centrals are similar to the total shown in Figure \ref{fig:gi_z008_dist_censat}, displaying functionally similar distributions.
The tensions on the location of the blue peak at the low-mass end (top row) and the late transition from blue- to red-dominated for \shark\trr from LC set \#1 are clearly visible for the centrals galaxies.

LC set \#2 produces centrals in better agreement with GAMA below $10^{10.5}M_\odot$ compared to LC set \#1, with the transition between blue/red populations more closely following that seen in GAMA.
While not shown in this work, we found this improvement to be primarily driven by the abundance matching and following stellar mass adjustment performed to generate LC set \#2.

Satellites show a markedly different distribution for our LC set \#1 at the lowest stellar mass bin, consistent with literature results that find satellites to be overly quenched \citep[e.g.,][]{weinmann2006,font2008,guo2016,cucciati2017}, displaying a distinctly redder distribution than GAMA.
By $M_\star\sim10^{10.0}M_\odot$, satellites do come quickly into a good agreement with the observations, both visually and according to our area matching criteria, displaying a notably better match.
Satellites incur a more drastic change from the change between sets \#1 and \#2, especially in the two lowest stellar mass bins, with a better agreement at $10^{9.7}M_\odot<M_\star<10^{9.9}M_\odot$ but a worse agreement at $10^{9.9}M_\odot<M_\star<10^{10.1}M_\odot$.
Unlike centrals, this improvement is not driven by the abundance matching, but from the central/satellite classification used.
The different classification is also responsible for the low-mass satellites distributions becoming bluer than GAMA, as this comes from contamination by centrals (by the classification in the simulation), which also show that trait.

\begin{figure}
    \centering
    \includegraphics[width=\linewidth]{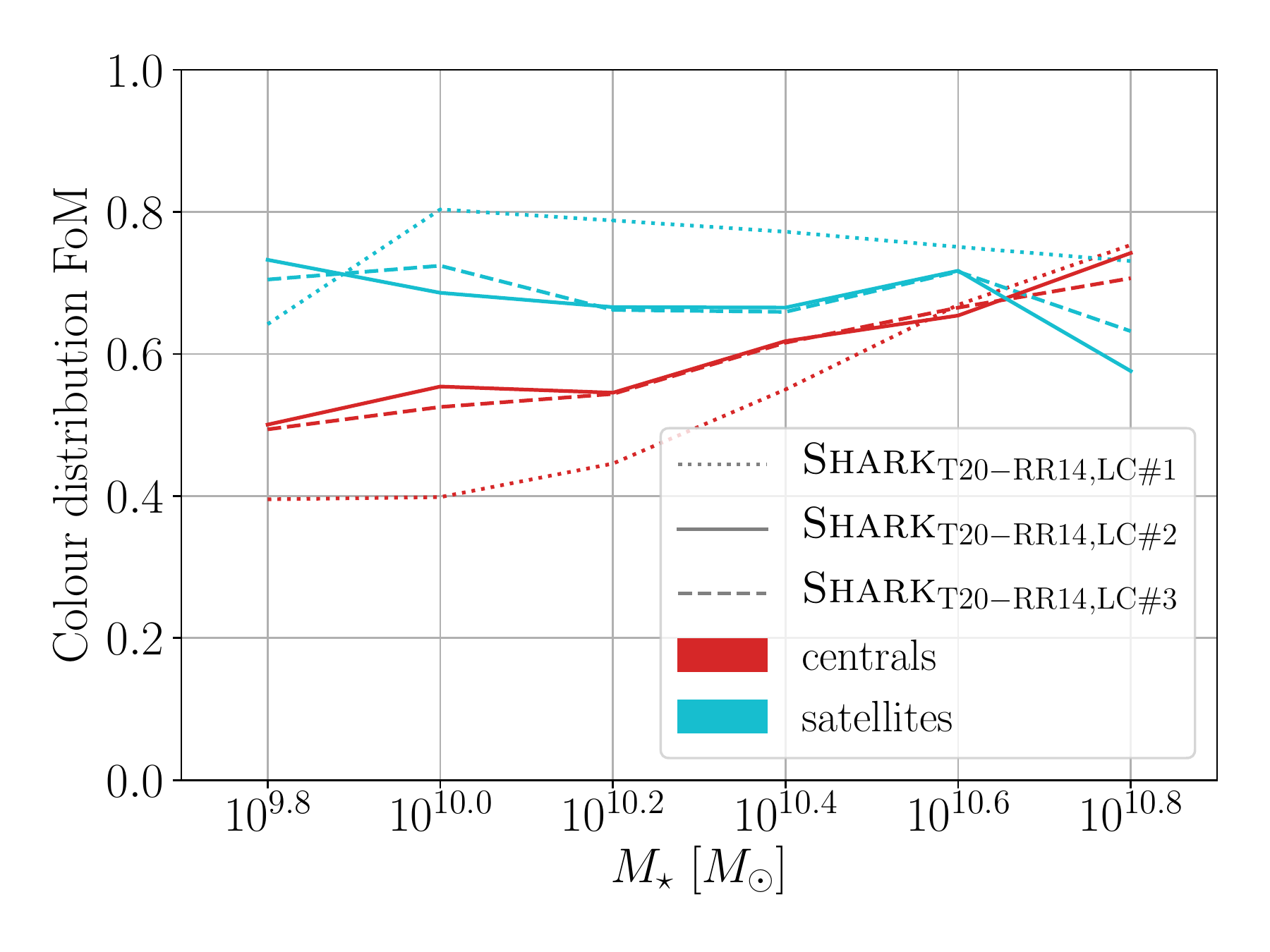}
    \caption{Colour distribution FoM of the colour distributions of \shark\trr\ from LC sets \#1, \#2 and \#3, shown in Figure \ref{fig:gi_z008_dist_censat}, as a function of stellar mass. FoM defined as the division of the area of the intersection between the photometry and GAMA, divided by the area of the union of both distributions. Line colours as in Figure \ref{fig:mock_nz_censat}.}
    \label{fig:gi_z008_match_censat}
\end{figure}

Figure \ref{fig:gi_z008_match_censat} shows the values from our FoM as a function of stellar mass.
The tension seen here at the low masses is driven by the excessively blue population for centrals.
As stellar mass increases, satellites quickly reach a very good agreement with GAMA, while centrals remain in larger disagreement due to the fraction of blue galaxies below $10^{10.7}M_\odot$ being too large.
It also shows the FoM for the \citetalias{robotham2011} group finder classification, where a noticeable improvement is seen in centrals below $10^{10.7}M_\odot$, and a similar performance above that stellar mass.
This comes at the cost of worsening the excellent match of satellites above $10^{9.9}M_\odot$, which should be expected as centrals dominate the number counts.
The change in stellar mass due to our abundance matching is the main driver for the improvement seen Figure \ref{fig:gi_z008_dist_all} from LC set \#1 to set \#2, and it is also the case for the improvement seen in centrals in \ref{fig:gi_z008_dist_censat}.
The improvement on the satellites in the $10^{9.9}$-$10^{10.1}M_\odot$ stellar mass bin is mainly driven by the use of the \citetalias{robotham2011} group finder.

Our method of randomly re-assigning $15\%$ of the centrals/satellites as satellites/centrals by construction produces a similar redshift distribution for each population to the one in GAMA. This percentage was chosen for exactly that reason, but it does not follow from this that one should expect an improved match in the distribution of the rest of the galaxy properties.
For that to be the case the differences between simulations and observations must be driven by classification errors by the group finding algorithm and not by the physics model.

The resulting colour distributions in Figure \ref{fig:gi_z008_dist_censat} and FoM in Figure \ref{fig:gi_z008_match_censat} show this to be the case, as these results closely mirror those form LC set \#2.
These results suggest that, while the use a group finder is the proper way to classify galaxies in a synthetic LC, the populations can be mimicked by a simple and inexpensive random reassignment of central/satellite status of a fraction of the galaxies, with $15\%$ providing a good match between \shark\ and GAMA.

This opens the possibility to use this same method to quickly reproduce central/satellite populations from \shark\ to other group catalogues, requiring only the use of a re-assignment percentage that mimics the expected classification confusion from the group catalogue to be compared to \shark. 
We warn the reader that this approach may not provide the desired results when using synthetic LC created with other methods, as the combination of \shark+\prospect\ is unique in its capability to provide a good match to observed colour distributions across a range of stellar masses and redshift, as shown in Appendix~\ref{sec:SAMcomp}.

From this comparison the main conclusion is that galaxies in \shark\ if taken directly from the simulation, the transition from blue- to red-dominated happens at too high stellar mass for centrals, and too low stellar mass for satellites, with centrals also being too blue before the transition. This tension in great part goes away when galaxies are classed centrals and satellites in the same way as done in observations.
From our analysis it is clear that the widely reported tension between simulations and observations, of satellites being overly quenched in models, may well be partly an artefact of the inherent limitation of group finders in observations.
In the next section we explicitly test this argument.
Note that the GAMA group finder already has a better purity than other group catalogues in part due to the high completeness of GAMA (see discussion in \citealt{robotham2011}).
This has to be carefully considered when comparing simulations with observations of satellites/central galaxies.


\subsection{Effect of classification on colour-derived red and passive fractions}\label{sec:passives}

The results from the previous section, while providing strong evidence that \shark+\prospect\ are capable of producing colour distributions similar to those observed, do not provide a straightforward comparison to the issues previous simulations have brought forward in the literature, commonly shown as the comparison of passive fractions using rest-frame colours.
To this end, we now compare \shark\trr\ to GAMA creating similar colour-based observational classifications of star-forming/passive galaxies found in the literature.
We do this to evaluate both the influence of how the central/satellite classification is performed on these measurements and the necessity for modified physical models.

For this, we have adapted two measurements.
The first one uses a single colour inspired by those used in \citet{weinmann2006,font2008}.
The second one uses a selection in colour-colour space similar to that used by \citet{williams2009}.
The colours used in this section, following the cited works, are in rest-frame absolute magnitudes.

\begin{figure}
    \centering
    \includegraphics[width=\linewidth]{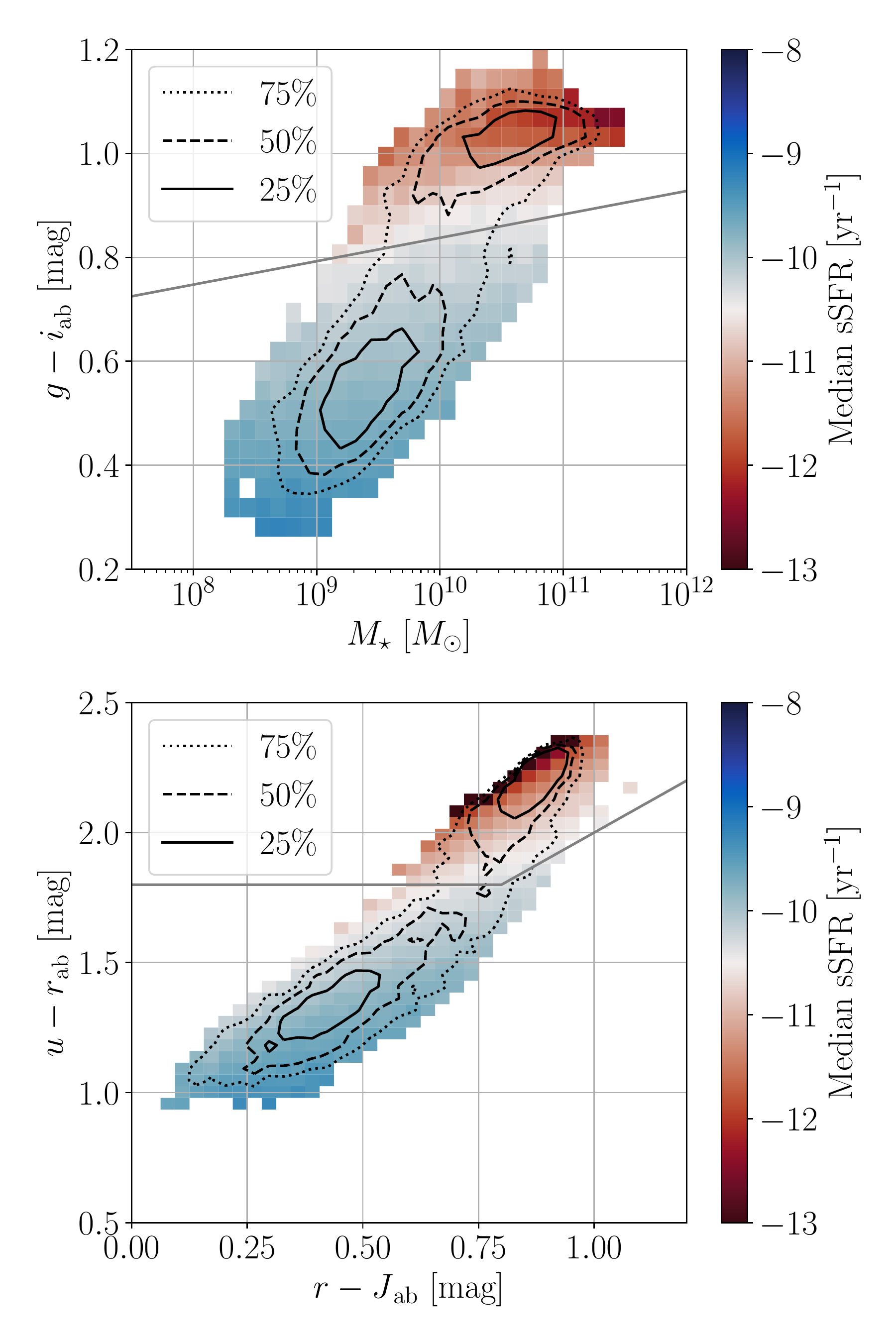}
    \caption{Colour classifications from GAMA. The top panel shows the absolute rest-frame colour distribution of galaxies in GAMA, as a function of stellar mass, coloured by the median sSFR of each bin. The black contours encircle the highest density regions containing 25\% (solid), 50\% (dashed) and 75\% of the galaxies. The colour map switches from red to blue at a sSFR of $10^{-10.5}$ yr$^{-1}$. The grey line shows the maximum $g-i$ value for a galaxy to be considered being blue. The bottom panel shows the absolute rest frame colour-colour distribution of galaxies in GAMA, coloured by the median sSFR of each bin. The black contours and colour map as in the top panel. Galaxies above the grey are considered as being passive.}
    \label{fig:colour_sel}
\end{figure}

Figure~\ref{fig:colour_sel} shows both our colour selections.
The top panel shows $g-i_\mathrm{ab}$ as a function of stellar mass, as found in GAMA, coloured by sSFR.
The choice for the middle point of the colour map to be at an sSFR of $10^{-10.5}$ yr$^{-1}$ is to visually separate the red sequence from the blue cloud.
From the dependence on both colour and stellar mass shown by the sSFR, we set the limit between blue and red galaxies at $(g-i)_\mathrm{ab}/\mathrm{mag}=0.05\log_{10}(M_\star/M_\odot)+0.35$, with galaxies above that line classified as red galaxies.

The bottom panel shows the $u-r_\mathrm{ab}$ versus $r-J_\mathrm{ab}$ distribution of galaxies in GAMA.
The bands used were chosen following the same argument as in \citet{williams2009}, that galaxies red in optical colours may well be dust-obscured star-forming galaxies, so the addition of a second colour that reaches into the infrared can serve to distinguish passive galaxies from dust-obscured ones.
The distribution we find in GAMA is similar to the one displayed in Figure~9 of \citet{williams2009}.
Since we are using a different filter set compared to their work, we choose to define our own selection criteria following the same principles, instead of performing a filter conversion and using the same limits they defined.
The colouring by the sSFR shows that this method of classification does indeed separate galaxies by star formation.
For galaxies bluer than $r-J_\mathrm{ab}=0.8$ we define galaxies redder than $u-r_\mathrm{ab}=1.8$ as passive, while for galaxies redder than $r-J_\mathrm{ab}=0.8$ we choose galaxies above the line defined by $u-r_\mathrm{ab}=r-J_\mathrm{ab}+1$
While this selection leaves out galaxies that we would classify as passive from a sSFR perspective, we decide against a more complex selection function as those are just a few galaxies, as seen from the contours in Figure \ref{fig:colour_sel}.

\begin{figure}
    \centering
    \includegraphics[width=\linewidth]{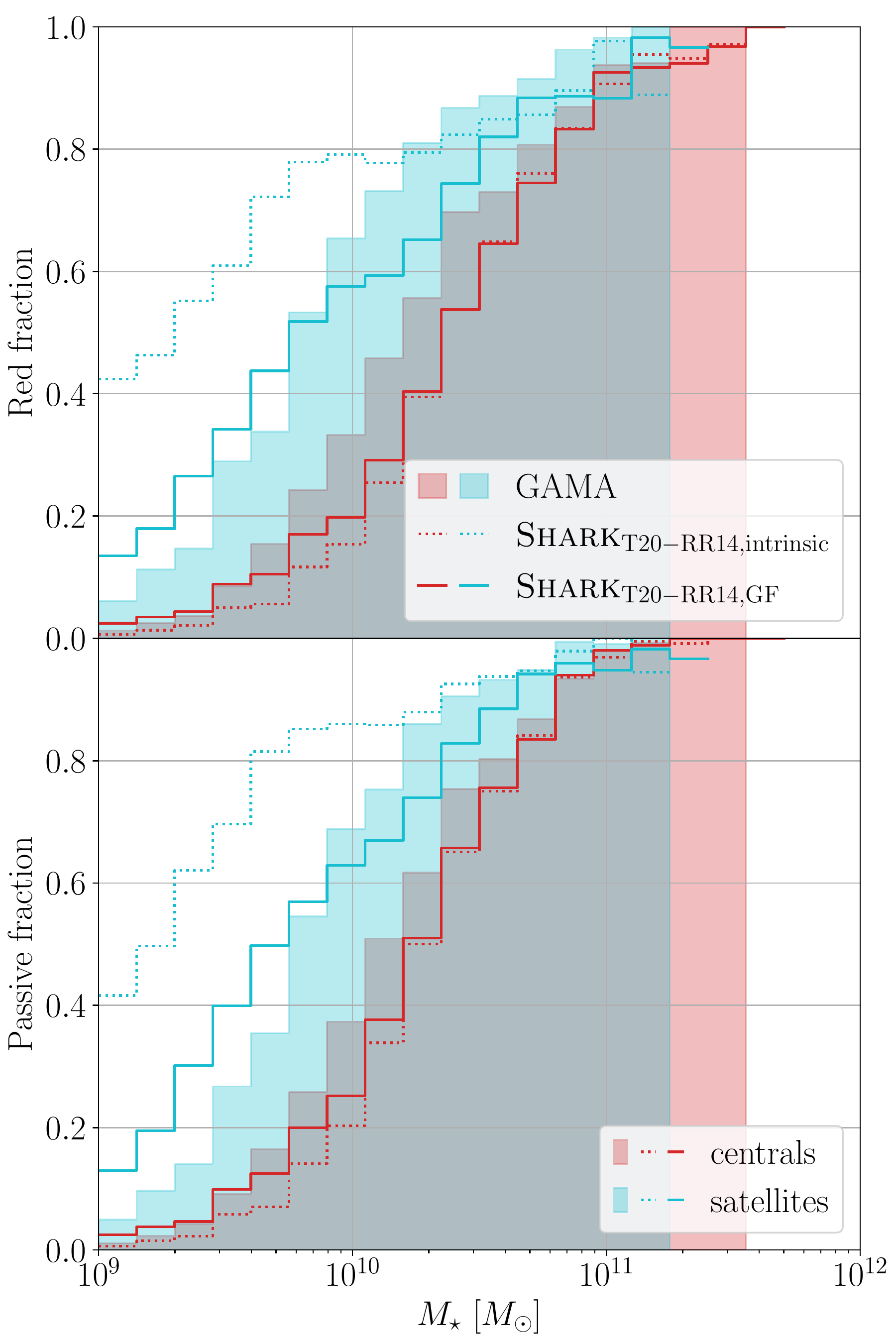}
    \caption{Comparison of colour selections between GAMA and \shark\trr\ (LC set \#2), for galaxies with $z<0.12$. The top panel shows the fraction of blue galaxies, as a function of stellar mass and galaxy type, divided by centrals and satellites. GAMA is shown by the solid histograms, \shark\trr\ with the central/satellite classification from the simulation and the solid lines for the \citetalias{robotham2011} group finder classification. The blue histograms show the fraction of blue satellites and the red ones for centrals. The bottom panel shows the fraction of passive galaxies in  GAMA and \shark\trr\ (LC set \#2), as a function of stellar mass and galaxy type, divided by centrals and satellites. Histogram colours and lines as in the top panel.}
    \label{fig:blue_passive}
\end{figure}

The result of applying these colour classifications to both \shark\trr\ and GAMA is shown in Figure~\ref{fig:blue_passive}, with the top panel showing the red fraction as a function of stellar mass, and the bottom the passive fraction as a function of stellar mass.
Going from intrinsic central/satellite classification to the \citetalias{robotham2011} group finder classification has a strong effect on the red and passive fractions of satellites.
Using the intrinsic one would lead us in the same direction as the work by \citet{font2008}, among others, that the physical modelling in \shark\ would be overly quenching satellite galaxies, but the switch to an observational classification almost completely solves that tension.
Some tension remains, even with the \citetalias{robotham2011} group finder classification, with satellites being still slightly too red below $\sim10^{10}M_\odot$, and becoming too blue above that.
Part of the decrease in red/passive fraction above $\sim10^{10}M_\odot$ stems from misclassified centrals, which display the same behaviour, irrespective of how they are classified as such. The latter is due to the sheer number of centrals being much larger than satellites.
Hence, a fraction of satellites being confused as centrals only barely impacts the central galaxy distribution.


\section{Discussion}\label{sec:discuss}

Our approach to reproduce observer-frame from simulations presented in this work is unprecedented in the literature.
When testing the predictions from simulations, it is common practice to choose single snapshots from a simulation that matches the redshift from the chosen observations for the comparison.
While time evolution would not be a critical factor for the redshift range on which we have focused on this work ($z<0.12$), using an LC instead of a snapshot box is critical for faithfully reproducing observations from a flux-limited survey (such as GAMA), especially when it comes to central/satellite classification.
Compared to our approach, most of the works in the literature directly compare to observations using the classification from the respective simulations \citet[e.g.,][]{henriques2015,guo2016,cora2018}.
The danger of that approach is that observational uncertainties can lead to misinterpretations of the results, leading to unwarranted changes to the physics modelling.
\citet{xie2020} show that different theoretical models predict different passive fractions of centrals/satellite galaxies.
Comparing these models with observations would be key to rule out some of these predictions.
However, we show here that systematic effects introduced by central/satellite confusion may be currently too large to be able to do this.

On that line, \citet{stevens2017} already pointed in that direction, showing that for \textsc{DarkSAGE} accounting for classification confusion in observations reduced the tension for the sSFR-derived passive fraction.
In this work, we provide strong evidence that a model that would be in tension with observations can be brought into a good agreement just by properly accounting for the limitations of observational catalogues.
The approach to this issue for the observations used by \citet{delucia2019} \citep[taken from][]{hirschmann2014}, of re-constructing the true passive fractions from the observed fractions and the uncertainty of the central/satellite classification \citep{weinmann2009}, may produce the same effect, but it is highly dependent on a good understanding of said uncertainties.
In contrast, our method, by construction, reproduces all shortcomings of observational classifications, as long as the simulation provides a good match to the properties relevant for the group finding.
For GAMA that means reproducing the observed $r$-band magnitudes, which we achieve through our abundance matching, and the spatial distribution of galaxies, which Appendix \ref{sec:GFcheck} shows is the case.

\begin{figure}
    \centering
    \includegraphics[width=\linewidth]{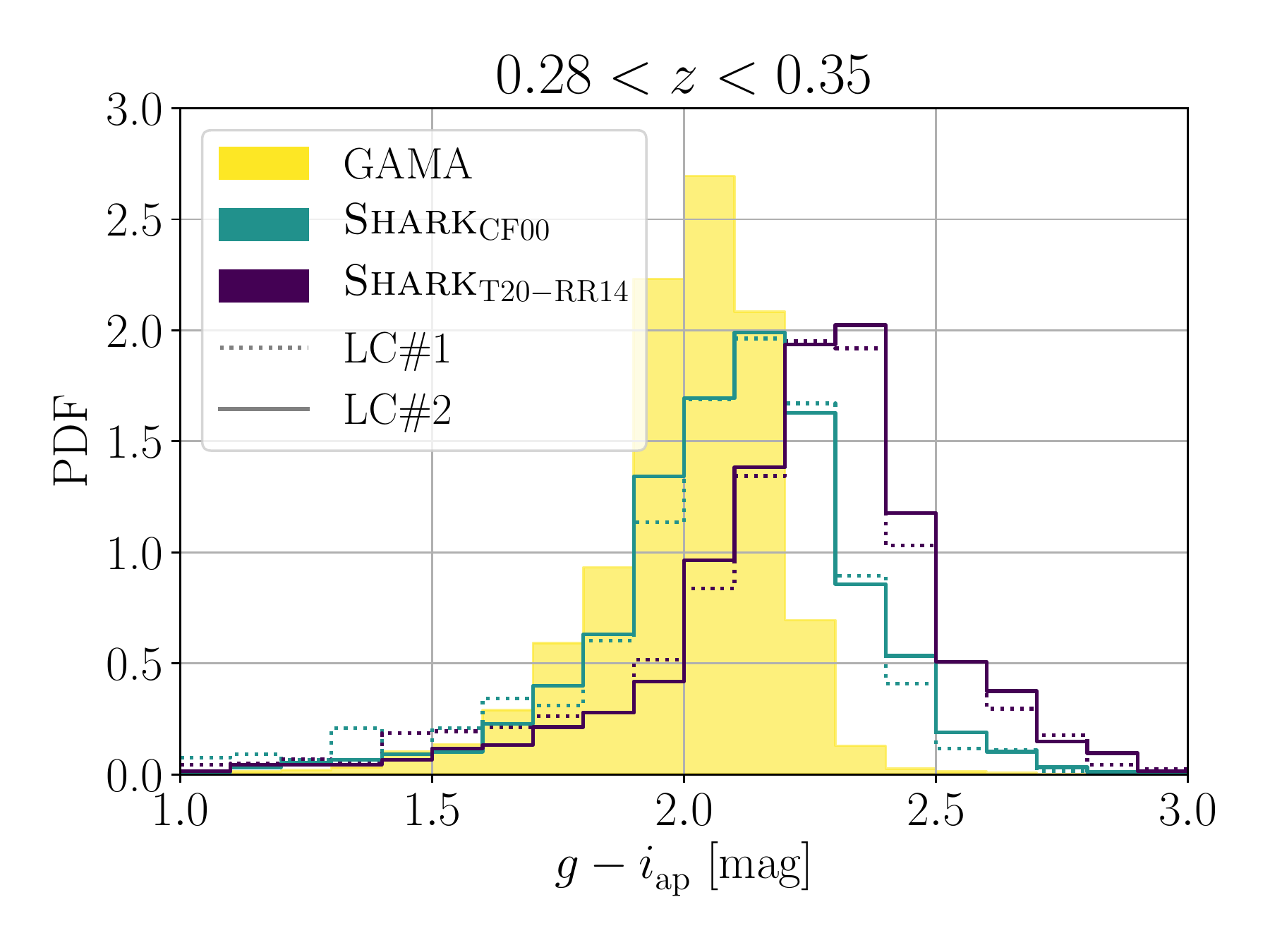}
    \caption{Apparent observer-frame $g-i$ colour distribution of galaxies with $0.28<z<0.35$ and $10^{11.3}M_\odot<M_\star<10^{11.5}M_\odot$, as in Figure \ref{fig:gi_z008_dist_all}, in LC sets \#1, \#2 and GAMA.}
    \label{fig:gi_z031_dist}
\end{figure}

While the attenuation model used in \prospect\ is not unique among SED fitting software, this approach is novel for forward modelling in simulations.
Appendix \ref{sec:SAMcomp} contains a comparison to two other SAMs, GALFORM and SAGE, for which the attenuation has been constructed using simpler models; applying a \citet{calzetti2000} extinction curve in the case of SAGE, or using idealised geometries for their radiative transfer calculation in the case of GALFORM.
From this comparison we suggest that \shark\trr, with its combination of \shark\, \prospect\ and the \citet{remy-ruyer2014}+\citet{trayford2020} models, provides a better representation of  the observations across a wide range of stellar masses and redshift.

One of the issues apparent with \shark\trr\ is that the colour distribution of galaxies below $\sim10^{10.5}M_\odot$ extends to bluer colours than seen in GAMA (Figure \ref{fig:gi_z008_dist_censat}).
While we focused on low-redshift galaxies for this work, at higher redshifts (where GAMA samples higher stellar masses) we find the opposite.
Figure \ref{fig:gi_z031_dist} shows that for massive galaxies, the colour distributions extend to redder colours than observed. This issue is related to the blue cloud being too blue at low redshift in \shark\trr.
Figure 15 in \citetalias{lagos2018} shows that the $Z_\mathrm{gas}$-$M_\star$ relation has a slope steeper than suggested in observations, meaning that for galaxies of low masses it will under-predict the amount of dust, while for massive galaxies the opposite will be true.
The consequence of this is seen in the colours of galaxies as discussed above.
Changes to the physics models in \shark\ that improves the $Z_\mathrm{gas}$-$M_\star$ relation without sacrificing agreement on other observables could provide an improved fit to the observed colour distributions.
Further changes are also required to solve the tension between \shark\ and GAMA at high redshift ($z\gtrsim0.4$); which goes from a strong tension on number counts seen in the LC set \#1 for the whole galaxy population, to a smaller but still significant tension on the number counts of satellites for LCs sets \#2 and \#3.
Abundance matching solves the former problem, but as centrals dominate the bulk number of galaxies, this has little effect on the satellites' contribution to the number of galaxies at high redshift.

In the near future, the Deep Extragalactic VIsible Legacy Survey \citep[DEVILS;][]{davies2018} will allow a significant extension of our analysis to $z\approx 1$.
DEVILS will deliver a catalogue of approximately 60,000 galaxies from $z\approx0.3$ to $z\approx1.0$ over an area of $\sim6\ \deg^2$, with a completeness $>90\%$.
These characteristics make DEVILS the ideal survey to study environmental effects over the last $8$~Gyr of Universe evolution.
This will allow us to replicate the tests done here but towards higher redshifts to identify new areas of agreement and tension.


\section{Conclusions}\label{sec:summary}

In this work we present a continuation of the exploration of the SED predictions from \shark\ combined with \prospect\ presented by \citetalias{lagos2019}.
Following the procedure of \citetalias{lagos2019} we constructed a set of LCs to simulate the GAMA survey.
We further refined these LCs by performing abundance matching, applying observationally-motivated errors, and using the same group finding algorithm as in GAMA (\citetalias{robotham2011}).
We compare the colour distributions from these synthetic LCs to the most recent catalogues available for GAMA \citep{robotham2011,liske2015,bellstedt2020a,bellstedt2020b}, finding that the default attenuation model adopted in \citetalias{lagos2019} provides a reasonable match to observations.

From these comparisons, it is clear that while it is the physics included in \shark\ what produces the colour bimodality, changes in the dust attenuation prescription modify both the peak and dispersion of both blue and red populations. We argue then that the choice of attenuation prescription is critical to reproduce the observed colour distributions.
Despite the success of \shark, some areas of tension remain in that the blue cloud of low-mass galaxies ($\lesssim10^{10}M_{\odot}$) tends to be $\approx0.1$ mag bluer than in GAMA, while the red sequence of massive ($\gtrsim10^{10.5}M_{\odot}$), intermediate redshift galaxies ($z\gtrsim0.3$) tend to be too red by $\approx 0.1$ mag.
These areas of tension are all related to the fact that \shark\ produces a gas metallicity-stellar mass relation that is steeper than observed, with low-mass galaxies ($\lesssim10^{10}M_{\odot}$) being slightly too metal poor, and massive galaxies ($\gtrsim10^{10.5}M_{\odot}$) being slightly too metal-rich.
Our study therefore suggests that a revision of the metal enrichment model of \shark\ is required in the future, but in a way that it does not compromise the overall success of the model.

We also analysed colour-derived red and passive fractions and find good agreement between \shark+\sting+\prospect\ and GAMA.
We find that reproducing the central/satellite classification from observations can solve tensions that otherwise appear for the colour distributions and red/passive fractions of satellites, which is a common issue on several galaxy formation models and simulations.
Hence, the long-standing problem of satellite galaxy over-quenching in galaxy formation simulations is at least in part an artefact of the limitations of group catalogues built in surveys such as GAMA.
Finally, we find that the effect of the classification used in GAMA can be reproduced by randomly re-assigning a fraction of centrals/satellites as satellites/centrals, with $15\%$ being the value required in \shark\ to match GAMA.
We caution though that other galaxy surveys with poorer completeness, such as SDSS, are likely to require a larger percentage of satellite/central contamination in order to mimic their limitations.


\section*{Acknowledgements}
We thank Chris Power and Pascal Elahi for their role in completing the SURFS $N$-body DM-only simulations suite, Rodrigo Tobar for his contributions to \shark, and the anonymous referee for their constructive report.
MB acknowledges the support of the University of Western Australia through a Scholarship for International Research Fees and Ad Hoc Postgraduate Scholarship.
CL is funded by the ARC Centre of Excellence for All Sky Astrophysics in 3 Dimensions (ASTRO 3D), through project number CE170100013.
CL also thanks the MERAC Foundation for a Postdoctoral Research Award. 
DO is a recipient of an Australian Research Council Future Fellowship (FT190100083) funded by the Australian Government.
SB acknowledges support by the \textit{Australian Research Council}'s funding scheme DP180103740.
This work was supported by resources provided by the Pawsey Supercomputing Centre with funding from the Australian Government and the Government of Western Australia.
The analysis on this work was performed using the programming language Python (\url{https://www.python.org}), and the open source libraries Matplotlib \citep{matplotlib}, NumPy \citep{numpy}, Pandas \citep{pandas} and SciPy \citep{scipy}.
Data used in this work was generated using Swinburne University's Theoretical Astrophysical Observatory (TAO). TAO is part of the Australian All-Sky Virtual Observatory (ASVO) and is freely accessible at \url{https://tao.asvo.org.au/tao/}.
The Millennium Simulation was carried out by the Virgo Supercomputing Consortium at the Computing Centre of the Max Plank Society in Garching. It is publicly available at \url{http://www.mpa-garching.mpg.de/Millennium/}.
The Semi-Analytic Galaxy Evolution (SAGE) model used in this work is a publicly available codebase that runs on the dark matter halo trees of a cosmological $N$-body simulation. It is available for download at \url{https://github.com/darrencroton/sage}.


\section*{Data availability}
The redshift \citep{liske2015}, \gc\ (\citetalias{robotham2011}) and random catalogue \citep{farrow2015} data is available as part of the GAMA Data Release 3 at \url{http://www.gama-survey.org/dr3/}.
The synthetic GAMA LC generated for this work will be shared on reasonable request to the corresponding author.
The new photometry and SED fitting data from GAMA were provided by Sabine Bellstedt by permission, and will be shared on request to the corresponding author with permission of Sabine Bellstedt.




\bibliographystyle{mn2e_trunc8}
\bibliography{papers}


\appendix

\section{Group finder quality check}\label{sec:GFcheck}

\begin{figure}
    \centering
    \includegraphics[width=\linewidth]{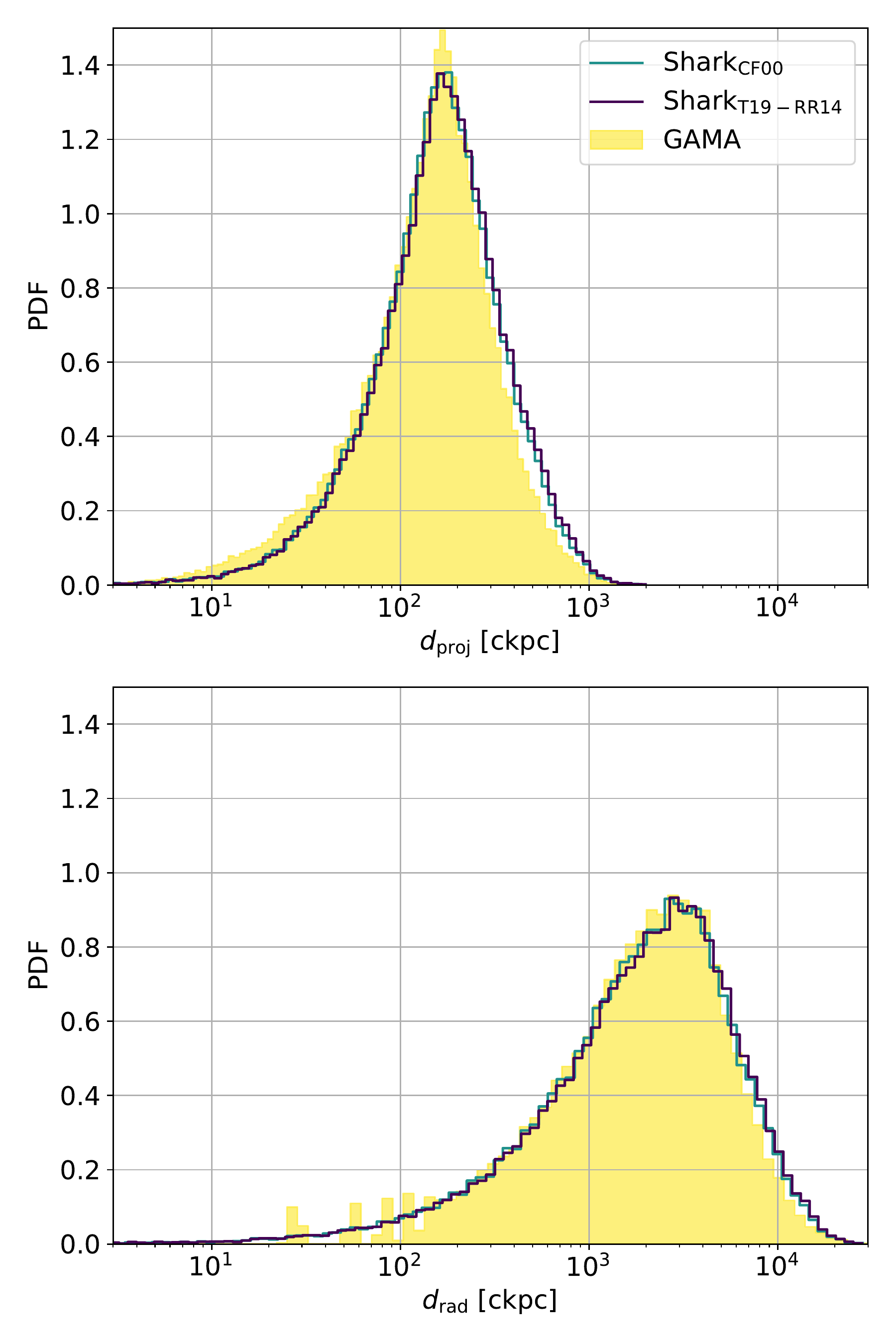}
    \caption{Comparison of the distribution of the radial and projected distance links generated by the \citetalias{robotham2011} group finder in both \gc\ and the synthetic LC set \#3.}
    \label{fig:links}
\end{figure}

To ensure that our use of the \citetalias{robotham2011} group finder produces sensible results, we checked several properties of the resulting group catalogues.
Figure~\ref{fig:links} shows the distribution of the most fundamental elements of this process, namely the links established between galaxy pairs, and compares them to those from \gc.
That the links created in our synthetic LC set \#3 closely follow the distribution observed in GAMA provides strong evidence that all the steps required to create our simulations were properly performed, and that we achieve a very good reproduction of observed galaxy distributions and properties.

\section{Comparison with other simulations}\label{sec:SAMcomp}

To gauge the achievements of \shark+\sting+\prospect\ in the broader context of theoretical galaxy formation models, we compare our results to two well-known SAMs, with their own methods to calculate SEDs: SAGE and GALFORM.
For SAGE we used the Theoretical Astrophysical Observatory \citep[TAO;][]{bernyk2016}, an established suite that combines one of several DM-only simulations with a choice of SAMs, IMFs and dust models, to produce a LC.
We produced a LC using the Millennium DM-only simulation \citep{springel2005}, a \citet{chabrier2003} IMF, the \citet{bruzual2003} stellar population synthesis libraries, and the \citet{calzetti2000} attenuation curve.
We have applied the GAMA selection of $r_\mathrm{ap}<19.8$, enforced a redshift limit of $0.0<z<0.6$, and given that TAO only handles single rectangular sky regions, an expanded selection centred on the G09 field with a similar footprint as GAMA ($\sim190\ \deg^2$). 
A summary of the choices made to produce this LC is presented in Table \ref{tab:STG_LC}.

\begin{table}
    \centering
    \begin{tabular}{c|c|c}
        \textbf{Simulation} & \textbf{SAM} & \textbf{Sky region}\\
        Millennium & SAGE & $129<\alpha<141$, $-8<\delta<8$\\
        \hline
        \textbf{Redshift} & \textbf{IMF} & \textbf{SED}\\
        $0.0<z<0.6$ & Chabrier & Bruzual \& Charlot\\
        \hline
        \textbf{Dust model} & \textbf{Selection}\\
        Calzetti & $r_\mathrm{ap}<19.8$
    \end{tabular}
    \caption{Summary of the properties of the SAGE LC, as how they are named on the submission form.}
    \label{tab:STG_LC}
\end{table}

The GALFORM LC used in this work is a pre-existing one, made using the version of the model from \citet{lagos2012}, constructing the LC following \citet{merson2013a}.
As with the SAGE LC, this was based on the Millennium simulation, though using a different set of merger trees \citep{jiang2014}.
GALFORM does the dust attenuation in a different way to that of SAGE and \shark\ and is described in detail in \citet{lacey2016}.
Briefly, dust is assumed to be in a two-phase medium, with birth clouds and a diffuse interstellar medium. 
The attenuation due to the diffuse medium is computed interpolating on a grid of radiative transfer calculations, while birth clouds are assumed to have a constant surface density and hence the attenuation is computed analytically.
The outcome of these processes is an optical depth that depends on galaxy properties galaxy size and gas mass and metallicity.
This attenuation model is applied in the LC used here.
This LC is noticeably smaller than the others, covering only the G09 equatorial field of GAMA ($129^o<\mathrm{RA}<141^o$, $-2^o<\mathrm{Dec}<3^o$).
This LC follows the same GAMA selection of $r_\mathrm{ap}<19.8$.

Before proceeding with the analysis, it is important to reiterate that the following results, and the underlying simulated galaxy SEDs, are a combination of the physical modelling each SAM uses and the respective prescriptions for stellar light emission, dust attenuation.
Building catalogues by fixing one of these components and varying the other would be highly informative on the choices made on each tool. For example, running \prospect\ on the outputs of each SAM would be a strong test of the modelling of physical processes of each.
Such a comparison, while highly desirable for the theoretical community, escapes the reach of this work, both by the complexity of modifying the outputs of the SAMs required by the SED modelling and the in-depth analysis needed to properly understand the differences.
For this reasons we approach this from an end-user perspective, where one would choose a set of tools that can produce the desired catalogues in a straightforward manner.

\begin{figure}
    \centering
    \includegraphics[width=\linewidth]{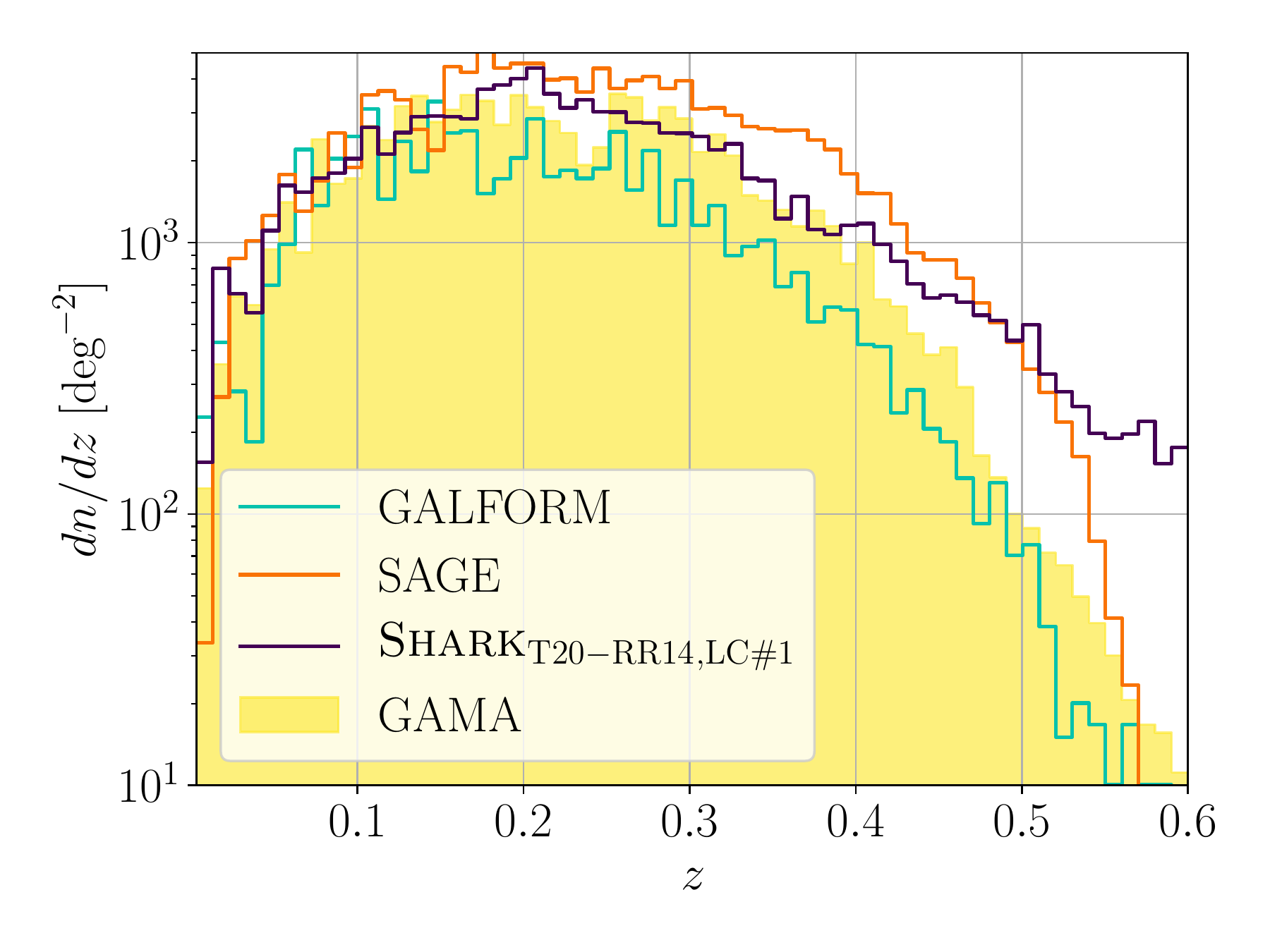}
    \caption{Redshift distribution of galaxies from GALFORM, SAGE \shark\trr\ LC set \#1 and GAMA. GAMA and \shark\trr as in Figure \ref{fig:mock_nz}. GALFORM is shown by the orange line and SAGE by the teal line.}
    \label{fig:SSG_nz}
\end{figure}

The redshift distributions for the three SAMs is shown in Figure \ref{fig:SSG_nz}.
While all are in good agreement with GAMA and each other for $z<0.2$, at higher redshifts the three show number counts that disagree with GAMA in different ways.
GALFORM slightly but consistently under-predicts the redshift distribution by a factor $\approx 2.5$. SAGE over-predicts the number counts by a factor of $\approx 2$ at $0.2 \lesssim z\lesssim 0.5$ to then sharply decline at $z\gtrsim 0.5$,  under-predicting the numbers of galaxies in the high-z tail. \shark\trr\ matches observations well at $z\lesssim 0.4$ and systematically deviates from observations at higher redshifts, predicting more galaxies than is observed. 

\begin{figure*}
    \centering
    \includegraphics[width=\linewidth]{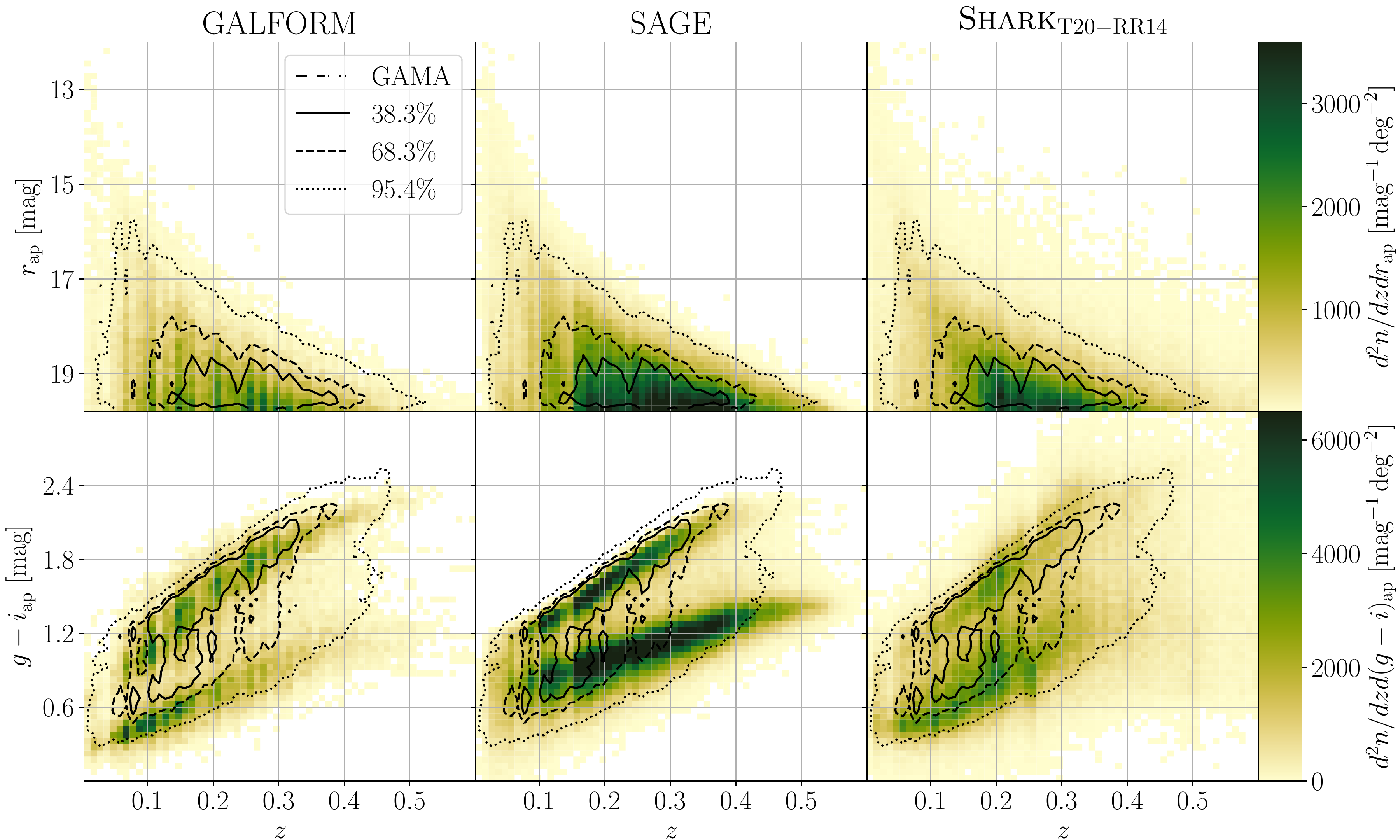}
    \caption{Magnitude (\rap) and colour (\giap) distributions from GALFORM, SAGE and \shark\trr\ compared to GAMA. The left column shows the distributions for GALFORM, the middle column for SAGE, and the right column for \shark\trr. Panel structure as in Figure \ref{fig:mock_prop}.}
    \label{fig:SSG_prop}
\end{figure*}

Figure~\ref{fig:SSG_prop} shows the magnitude and colour distributions of the three, as a function of redshift.
All three produce sensible red populations, with the largest differences being on the blue population.
Both GALFORM and SAGE produce distinct, narrow blue populations. GALFORM  consistently produces a blue cloud that is too blue compared to GAMA. SAGE on the other hand, has a blue cloud that extends to higher redshifts than observed and is too shallow compared to GAMA.
\shark\trr instead produces sensible a sensible blue cloud that has a better mode, scatter and slope. 

\begin{figure*}
    \centering
    \includegraphics[width=\linewidth]{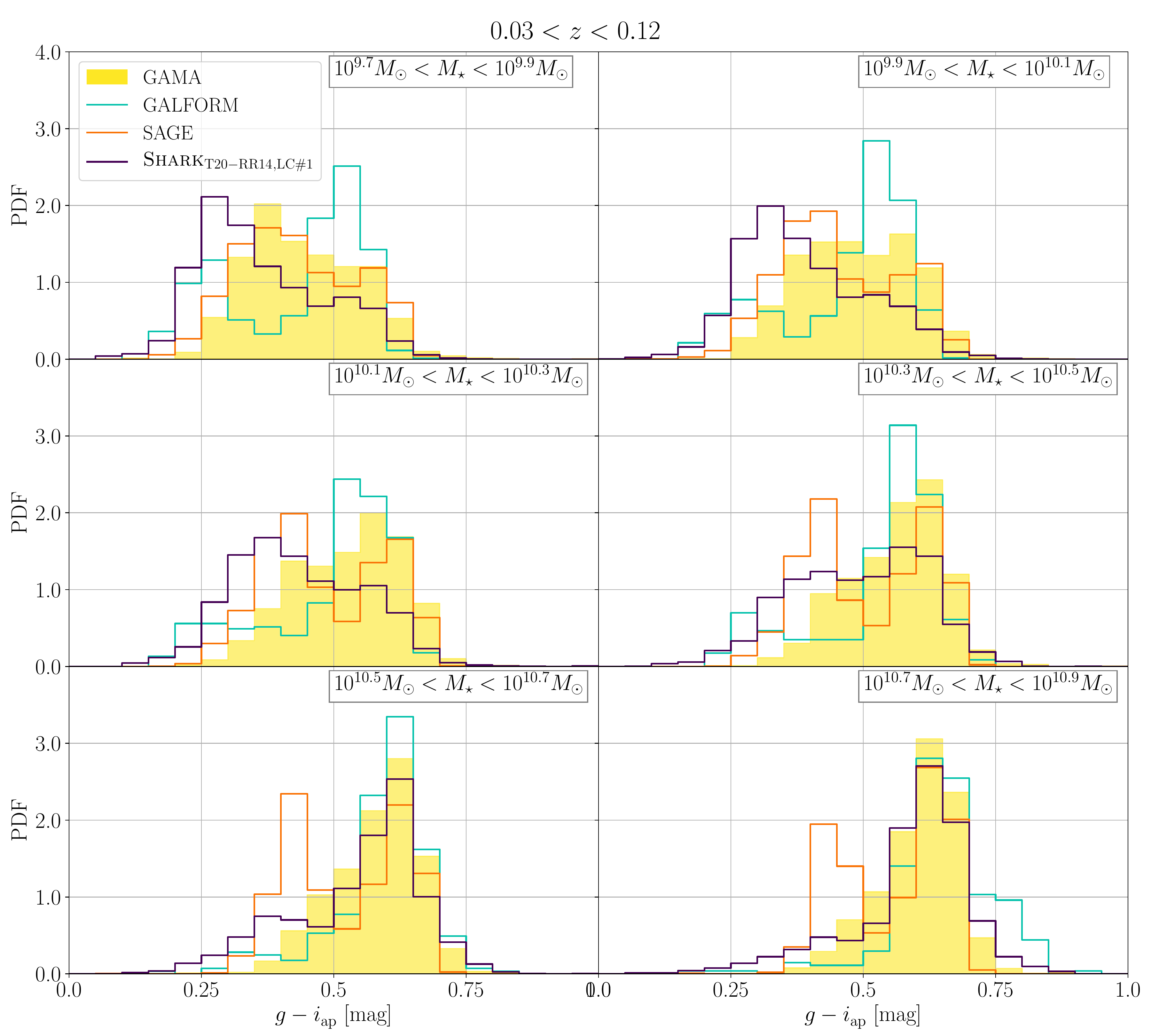}
    \caption{Apparent observer-frame $g-i$ colour distribution of galaxies with $z<0.12$ \shark\trr\ from LC set \#1, GALFORM and SAGE and GAMA. The figure structure is the same as Figure \ref{fig:gi_z008_dist_all}, and colours as in Figure \ref{fig:SSG_nz}.}
    \label{fig:SSG_gi_dist}
\end{figure*}

Figure \ref{fig:SSG_gi_dist} shows a comparison between the colour distributions of GAMA with \shark\trr, SAGE and GALFORM.
Both SAGE and GALFORM display unrealistically bimodal colour distributions, though on different ends of the stellar mass range.
SAGE produces a good fit for the observed distribution at the lowest mass bin, but as stellar mass increases the number of galaxies with ${g-i}_\mathrm{ap}\sim1.0$ quickly diminishes, splitting the distribution into two.
While the red population is in decent-to-good agreement with GAMA, the blue population comes into clear tension by $M_\star\sim10^{10.4}M_\odot$.
GALFORM on the other hand noticeably under-predicts the fraction of galaxies with ${g-i}_\mathrm{ap}\sim0.75$ for stellar masses below $\sim10^{10.4}M_\odot$.


\bsp	
\label{lastpage}
\end{document}